\documentclass[useAMS,usenatbib]{mn2e}
\usepackage{graphicx,amsmath,color,amssymb}
\usepackage[pdftitle={Massive Neutrinos and the Non-linear Matter Power Spectrum}]{hyperref}
\usepackage[all]{hypcap}

\newcommand{\adsurl}[1]{\href{#1}{ADS}}

\newlength{\narrowfigurewidth}
\newlength{\figurewidth}
\newlength{\widefigurewidth}
\newcommand{\etal}
  {et al.}

\newcommand{\Lya}{Lyman-$\alpha\;$}

\newcommand{\Mpch}{\,\mathrm{Mpc} \,h^{-1}}
\newcommand{\hMpc}{\,h \mathrm{Mpc}^{-1}}
\newcommand{\gadget}{{\small GADGET\,}}
\newcommand{\halofit}{{\small HALOFIT\,}}

\title[The Matter Power with Neutrinos]{Massive Neutrinos and the Non-linear Matter Power Spectrum }

\author
  [S. Bird \etal]
  {Simeon Bird $^{1}$\thanks{E-mail: spb41@ast.cam.ac.uk},
  Matteo Viel $^{2,3}$\thanks{E-mail: viel@oats.inaf.it} and 
  Martin G. Haehnelt $^{1}$\thanks{E-mail: haehnelt@ast.cam.ac.uk}
\vspace{7mm}\\
$^1$Institute of Astronomy and Kavli Institute for Cosmology, Madingley Road, Cambridge CB3 0HA, U.K.\\
$^2$INAF - Osservatorio Astronomico di Trieste, Via G.B. Tiepolo 11, I-34131 Trieste, Italy \\
$^3$INFN/National Institute for Nuclear Physics, Via Valerio 2, I-34127 Trieste, Italy
}

\begin{document}

\pagenumbering{alph}
\date{}

\maketitle
\pagerange{\pageref{firstpage}--\pageref{lastpage}} \pubyear{2011}

\pagenumbering{arabic}
\label{firstpage}

\begin{abstract}
We perform an extensive suite of N-body simulations of the matter
power spectrum, incorporating massive neutrinos in the range 
$M_\nu = 0.15 - 0.6$~eV, probing the non-linear regime at
scales $k<10 \hMpc$ at $z<3$.  We extend the widely used 
\halofit~approximation to account 
for the effect  of  massive neutrinos on the power spectrum. 
In the strongly non-linear regime  \halofit~systematically 
over-predicts the suppression due to the free-streaming
of the neutrinos. The maximal discrepancy occurs at $k \sim 1 \hMpc$, and
is at the level of $10$\% of the total suppression.
Most published constraints on neutrino masses based 
on \halofit are not affected, as they rely on
data probing the matter power spectrum in the linear or mildly
non-linear regime. However, predictions for future galaxy, 
Lyman-$\alpha$ forest and weak lensing surveys extending 
to more non-linear scales will benefit from the improved
approximation to the non-linear matter power spectrum we provide.
Our approximation reproduces the induced neutrino suppression 
over the targeted scales and redshifts significantly
better. We test its robustness with regard to changing 
cosmological parameters and a variety of modelling effects.
\end{abstract}

\begin{keywords}
        neutrinos - cosmology: large-scale structure of Universe - cosmology: dark matter
\end{keywords}
 
\section{Introduction}
\label{sec:intro}

A variety of solar, atmospheric, reactor and accelerator
neutrino experiments have firmly established the existence of neutrino 
flavour oscillations,  implying that neutrinos have at least two
non-zero mass eigenstates \citep{Becker-Szendy:1992, Fukuda:1998, Ahmed:2004}.
These experiments yield a lower limit on the sum of the neutrino 
masses; $M_\nu > 0.05 eV$. Classical $\beta$-decay experiments 
provide an upper limit on the mass of the electron 
neutrino of $2.2$~eV \citep{Lobashev:2003}, 
but may in future reach constraints as low as $0.2$~eV \citep{Wolf:2010}. 

Sensitivity to well below $1$~eV is necessary to distinguish the neutrino 
mass hierarchy \citep{Jimenez:2010}, which is of
particular interest when formulating theories of particle physics beyond 
the standard model.  Fortunately, such strong constraints can be obtained 
from the cosmic neutrino background. Simply requiring $\Omega_\nu
\leq \Omega_M \approx 0.3$ and fixing the Hubble constant  
implies $M_\nu \lesssim 15$~eV  \citep{Gershtein:1966,
 Lesgourgues:2006}.  The cosmic microwave background (CMB; e.g. \cite{WMAP7}) 
combined with measurements of the matter power spectrum from galaxy
surveys provide more stringent  constraints, presently around $M_\nu \lesssim 0.6$~eV 
\citep{Elgaroy:2002,Seljak:2006,Tegmark:2006,Gratton:2007,Mantz:2009,Thomas:2009,Reid:2010}.
\cite{Seljak:2006} combined the \Lya forest flux power spectrum with the CMB
to claim the tightest constraint so far; $M_\nu<0.17$~eV.
This strong limit may, however, be partly caused by a
slight tension between the power spectrum amplitude preferred by the
\Lya forest data and that preferred by the CMB. In any case, it should be     
tested and improved by upcoming data \citep{Slosar:2011}.
The \Lya data alone provides  a limit 
of $M_\nu < 1$~eV \citep{Viel:2005,Viel:2010}, slightly stronger than
that provided by the CMB data alone \citep{Larson:2011}.

The sensitivity of the matter power spectrum to neutrino mass is due
to the free-streaming of the neutrinos which  suppresses 
the growth of structure. On large scales, where matter 
perturbations can be described by linear theory, 
this effect is relatively straightforward to quantify
with numerical simulations or approximate schemes 
(e.g. \cite{Bond:1980,Zeldovich:1980, Ma:1993, Valdarnini:1998}) 
using standard Boltzmann codes \citep{CAMB, Lesgourgues:2011}. 
Probing the effect of neutrinos on smaller, more non-linear scales
is more challenging, but has been attempted by, for example, 
\cite{Saito:2008, Saito:2009, Saito:2011}, \cite{Wong:2008} and \cite{Lesgourgues:2009},
 with extensions to perturbation 
theory. Also of interest here is the approach of \cite{Abazajian:2005, Hannestad:2005}, who
described the fully non-linear regime by perturbatively evaluating the effect
of massive neutrinos on halo profiles. 

However, accurately probing the effect of massive neutrinos on the growth 
of structure arbitrarily far into the non-linear regime requires
fitting formulae precisely calibrated from high dynamic range N-body simulations. 
The formula we present here builds on
the widely-used \halofit~\citep{Smith:2003}.
\halofit~models the non-linear matter power spectrum by assuming that
the growth of structure is due to a combination of the 
density structure of individual DM haloes and the clustering of these
DM haloes relative to each other. It has been calibrated against 
a wide range of CDM simulations, superseding earlier prescriptions based on the
stable clustering hypothesis \citep{Hamilton:1991, Peacock:1996}.
At low and moderate redshifts \halofit~models the non-linear growth of
structure in a $\Lambda$CDM cosmology
with an accuracy of $\lesssim 5-10$\% \citep{Heitmann:2010a} for $k < 1 \hMpc$, 
but it has not been calibrated for non-zero neutrino masses. 

Fortunately, the quality of the numerical simulations
has advanced to the point where it is possible
to model the impact of neutrinos to the level of a few percent, 
despite the difficulties presented by the large thermal dispersion of light 
neutrinos \citep{Brandbyge:2008, Viel:2010}. 
A quantitative  numerical study of the suppression of the matter power
spectrum due to free-streaming of neutrinos appears timely, as future matter 
power spectrum surveys aim to  make measurements   
accurate to a few percent. Realising the potential of 
these measurements to constrain the neutrino mass 
requires tight control of systematic error from
theoretical predictions.

We present results from simulations  with the recently developed version of the 
Tree-SPH code \gadget-3 \citep{Springel:2005,Viel:2010}. This suite of
simulations was designed to evaluate the performance of
\halofit~predictions when compared to 
full numerical simulations. We use our simulations to 
calibrate an improved fitting formula, capable of accurately reproducing the 
effect of massive neutrinos on the non-linear matter power spectrum. 
We discuss our methods in Section \ref{sec:methods}, 
present our results in Section \ref{sec:results}, and 
conclude in Section \ref{sec:discussion}.

\section{Modelling the non-linear matter power spectrum}
\label{sec:methods}

\begin{figure}
\includegraphics[width=0.45\textwidth]{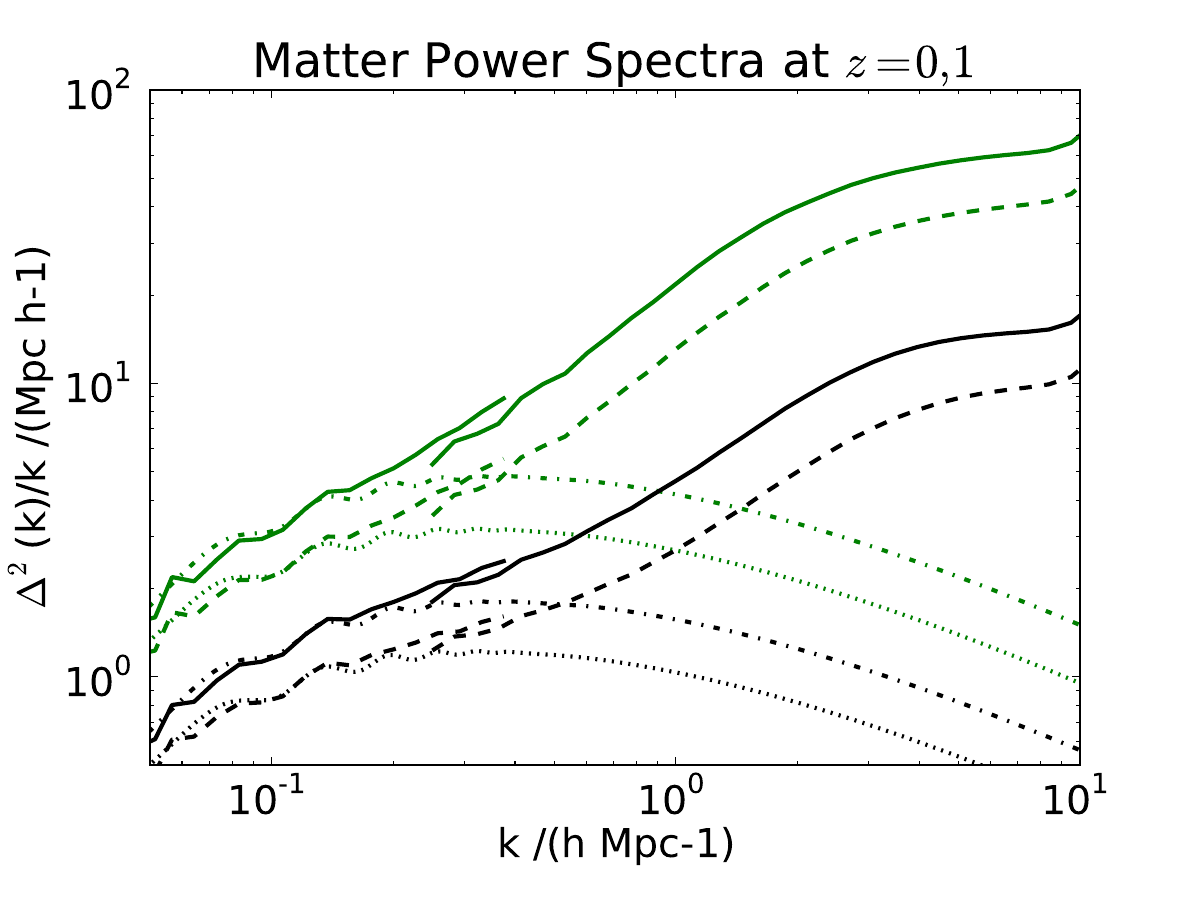}
\caption{Matter power spectra from a simulation with massive neutrinos (dashed lines), and $M_\nu = 0.6$~eV, 
compared to the corresponding power spectra from a $\Lambda$CDM simulation (solid lines). 
Dotted lines show the linear theory prediction for massive neutrinos, while dot-dashed lines
show the linear theory prediction for $\Lambda$CDM. 
Upper (green) lines show spectra at $z=0$, while lower (black) lines show them at $z=1$. 
The simulations shown are S60 (small scales) and L60 (large scales), 
whose parameters are described in Table \ref{tab:partsimuls}.
S60 shows only modes where $k > 4 \times 2\pi /$ Box-size, so sample variance is small, 
and L60 is cut off at $k= 0.4 \hMpc$, for clarity. Note that the absolute
power spectrum is slightly less converged with respect to box size than 
the ratio of the massive neutrino power spectrum with and without massive neutrinos.
}\label{fig:absolute}
\end{figure}

Here we briefly describe the numerical methods we use to simulate
structure growth in the presence of 
massive neutrinos, as well as  the approximations underlying  \halofit. 
In Section \ref{sec:neutrinogrowth} we review the predicted effect of neutrinos in linear theory.
In Section \ref{sec:simulations} we discuss the strategies used for simulating 
the effect of massive neutrinos, describing methods based on both particle and 
fourier-space representations of neutrinos. 
In Section \ref{sec:halofit}, we summarise \halofit. 
Finally, Section \ref{sec:simdetails} gives details of our numerical simulations.

\subsection{Massive neutrinos in linear theory}
\label{sec:neutrinogrowth}

Neutrinos in the very early Universe are relativistic and tightly 
coupled to the primordial plasma. They decouple while still relativistic and 
then redshift adiabatically. 
The transition to non-relativistic behaviour takes
place when $M_\nu = 3T_\nu$, at a redshift of $(1+z) = 2 \times 10^3 (M_\nu/ 1\,\mathrm{eV})$. 
However, it is some time later before relativistic corrections are negligible, especially in 
the high-velocity tail of the Fermi-Dirac distribution.
Once non-relativistic, massive, weakly-interacting neutrinos behave qualitatively 
as a species of warm/hot dark matter, suppressing 
fluctuations on scales smaller than their thermal free-streaming length. 

This suppression has a distinctive shape in the linear regime. Neutrinos of mass $M_\nu$ have energy density 
$\Omega_\nu$, and make up a fraction, $f_\nu$, of the dark matter, where
\begin{align}
        &\Omega_\nu = \frac{M_\nu}{93.14 \,h^2 \mathrm{eV}}\,,&f_\nu = \frac{\Omega_\nu}{\Omega_M}\,.
        \label{eq:neutdensity}
\end{align}
$\Omega_M$ is the present day matter density and $h$ is the Hubble constant. 
The residual comoving peculiar thermal velocity of the neutrinos at redshift $z$ is
\begin{equation}
        v_\mathrm{th} \approx 150 (1+z) \left[ \frac{1 \,\mathrm{eV}}{M_\nu}\right]\, \mathrm{km /s }\,.
        \label{eq:therm}
\end{equation}
By analogy with the Jeans length, we can define a co-moving free-streaming scale
\begin{equation}
        k_\mathrm{FS} = \sqrt{\frac{2}{3}}\frac{H(t) a(t)}{v_\mathrm{th}(t)}\,.
        \label{eq:freestream}
\end{equation}
Neutrinos cannot cluster on scales smaller than $k_\mathrm{FS}$, 
and so the matter power spectrum, defined as 
\begin{equation}
        P(k) = | \tilde{\delta}(k) |^2\,,
        \label{eq:powerspec}
\end{equation}
where a tilde denotes a Fourier transformed quantity, 
 is suppressed by a constant factor proportional 
to the neutrino energy density, $f_\nu$. The Boltzmann equation 
can be solved numerically to show that the fractional 
change in the power spectrum for $f_\nu < 0.07$ is
\begin{equation}
        \frac{\delta P}{P} \approx -8 f_\nu\,.
        \label{eq:suppression}
\end{equation}
On larger scales, $k < k_\mathrm{FS}$, 
neutrinos have free-streamed only for a portion of the cosmic history, 
inducing less suppression. The largest scale on which 
neutrinos alter the matter power spectrum is given by the wavenumber for 
which they free-streamed only immediately after becoming non-relativistic \citep{Lesgourgues:2006}
\begin{equation}
        k_\mathrm{nr} \sim 0.018 \sqrt{\Omega_M \left[ \frac{M_\nu}{1 \,\mathrm{eV}}\right]} \hMpc\,.
\end{equation}
Today, neutrinos in the observationally acceptable range have 
$k_\mathrm{nr} \sim 10^{-3}$, placing the free-streaming effect at the 
scales probed by galaxy and \Lya measurements of the matter power spectrum, as illustrated by  
Figure \ref{fig:absolute}, which shows non-linear matter power spectra estimated from our simulations.
For further information on the effect of neutrinos in linear theory see, e.g.
\cite{Kolb:1990, Bond:1980, Eisenstein:1999} and references therein. 

\subsection{Incorporating massive neutrinos into N-body simulations}
\label{sec:simulations}

The first N-body simulations incorporating massive neutrinos were aimed at 
evaluating their potential as a dark matter candidate \citep{White:1983,Klypin:1992,Ma:1993}.
Recently attention has shifted to their perturbative effect as a sub-dominant component 
in a mainly $\Lambda$CDM universe.
We build on the work of \cite{Viel:2010}, who incorporated massive neutrinos 
into \gadget-3 \citep{Springel:2005} in two ways; as a new particle species, 
and using a Fourier-space representation \citep{Brandbyge:2009}. 
\cite{Viel:2010} have described the implementation in detail and
performed extensive validation tests. 
Here we briefly review the two implementations of neutrinos. 
We refer the reader  to \cite{Brandbyge:2008, Brandbyge:2009} and \cite{Viel:2010} 
for further details. Both the particle and the Fourier space
implementations incorporate massive neutrinos 
directly into the simulations and so are significantly more accurate than 
simulations altering only the initial conditions of a $\Lambda$CDM 
simulation, as {\it e.g.} in \cite{Agarwal:2011}.

All our simulations are inherently Newtonian, ignoring relativistic corrections.
It is therefore important to set up the initial conditions of the simulation 
at sufficiently low redshift that the vast majority of neutrinos are non-relativistic. 
Otherwise, neutrinos will be treated as massive while still effectively massless, over-estimating 
their effect by a constant factor in the linear regime, at the highest redshifts.

\subsubsection{Particle-based Neutrinos}

Massive neutrinos can be incorporated into cold dark matter simulations as a separate
low-mass collisionless particle species. Structure growth is primarily driven by 
the cold dark matter. The neutrino species has significant thermal dispersion, preventing it from
clustering on small scales. This allows us to avoid following the small-scale evolution of the neutrino 
component, first by ignoring the short-range tree force for the neutrinos and 
second by setting the timestep purely from the dark matter species, effectively relaxing 
the Courant conditions for the neutrino component.
\cite{Brandbyge:2008} and \cite{Viel:2010} found that these approximations
significantly speed up computation while having a negligible impact on results.

One potential problem with particle based simulations is thermal shot noise from the 
finite sampling of the neutrino distribution \citep{Wang:2007}, especially for the higher
thermal velocities encountered at high redshift or for particularly low mass neutrinos. 
We deal with this by explicitly checking the effect of shot noise, and increasing the
neutrino particle number should it be non-negligible.

\subsubsection{Fourier-space Neutrinos}

The effect of neutrinos on the dark matter can also be approximated by adding their 
gravitational force to that of the dark matter in Fourier space \citep{Brandbyge:2009}, 
as part of the particle mesh computation. The massive neutrino component
is assumed to have the power spectrum predicted by linear theory, in our case computed 
using the Boltzmann code CAMB\footnote{\url{http://www.camb.info}} \citep{CAMB}.
Since the neutrinos are not followed directly, we can avoid shot noise and 
the need to deal with large thermal velocities when computing their 
effect on the overall growth of structure.
However, because the neutrinos are evolved using linear theory, Fourier-space simulations 
neglect any non-linear evolution in the neutrino component itself, including back-reaction 
from non-linearities in the dark matter. As we will discuss in Section \ref{sec:results} 
these effects are important, and so we shall focus mainly on simulations involving particles. 

\begin{table*}
\begin{center}
\begin{tabular}{|c|c|c|c|c|c|c|c|c|c|c|c|c|}
\hline
Name   & $M_\nu$ (eV) & Box ($\Mpch$) & $N_\mathrm{DM}^{1/3}$ & $N_\mathrm{bar}^{1/3}$ & $N_\nu^{1/3}$ & No. Seeds & $z_i$ & $\Omega_M$ & $A_\mathrm{s}/10^{-9}$ & $n_s$ & $h$ & $\sigma_8$ (derived) \\
\hline 
S15   & $0.15$ & $150$ & $512$ & $0$    & $512$  & $6$ & $24$ & $0.3$  & $2.27$ & $1.0$ & $0.7$  & $0.84$ \\
S30   & $0.3$  & $150$ & $512$ & $0$    & $512$  & $6$ & $49$ & $0.3$  & $2.27$ & $1.0$ & $0.7$  & $0.80$ \\
S60   & $0.6$  & $150$ & $512$ & $0$    & $512$  & $6$ & $99$ & $0.3$  & $2.27$ & $1.0$ & $0.7$  & $0.73$ \\
L15   & $0.15$ & $512$ & $512$ & $0$    & $512$  & $1$ & $24$ & $0.3$  & $2.27$ & $1.0$ & $0.7$  & $0.84$ \\
L30   & $0.3$  & $512$ & $512$ & $0$    & $512$  & $1$ & $49$ & $0.3$  & $2.27$ & $1.0$ & $0.7$  & $0.80$ \\
L60   & $0.6$  & $512$ & $512$ & $0$    & $512$  & $1$ & $99$ & $0.3$  & $2.27$ & $1.0$ & $0.7$  & $0.73$ \\
S15NU & $0.15$ & $150$ & $512$ & $0$    & $1024$ & $1$ & $24$ & $0.3$  & $2.27$ & $1.0$ & $0.7$  & $0.84$ \\
S60NU & $0.6$  & $150$ & $512$ & $0$    & $1024$ & $1$ & $99$ & $0.3$  & $2.27$ & $1.0$ & $0.7$  & $0.73$ \\
S60AS & $0.6$  & $150$ & $512$ & $0$    & $512$  & $1$ & $99$ & $0.3$  & $2.0$  & $1.0$ & $0.7$  & $0.73$ \\
S60NS & $0.6$  & $150$ & $512$ & $0$    & $512$  & $1$ & $99$ & $0.3$  & $2.27$ & $0.9$ & $0.7$  & $0.73$ \\
S60ND & $0.6$  & $150$ & $640$ & $0$    & $640$  & $1$ & $99$ & $0.3$  & $2.27$ & $1.0$ & $0.7$  & $0.73$ \\
S60H  & $0.6$  & $150$ & $512$ & $0$    & $512$  & $1$ & $99$ & $0.3$  & $2.27$ & $1.0$ & $0.75$ & $0.79$ \\
S30OM & $0.3$  & $150$ & $512$ & $0$    & $512$  & $2$ & $49$ & $0.25$ & $2.27$ & $1.0$ & $0.7$  & $0.68$ \\
S30OL & $0.3$  & $150$ & $512$ & $0$    & $512$  & $1$ & $49$ & $0.4$  & $2.27$ & $1.0$ & $0.7$  & $1.07$ \\
S60BA & $0.6$  & $150$ & $512$ & $512$  & $512$  & $1$ & $99$ & $0.3$  & $2.27$ & $1.0$ & $0.7$  & $0.73$ \\
\hline
\end{tabular}
\end{center} 
\caption{Summary of simulation parameters. $\sigma_8$ has been derived
  from the other parameters. For each simulation with massive
  neutrinos we ran a $\Lambda$CDM simulation with identical
  parameters, but without massive neutrinos, in order to compute their
  ratio. The value of $\sigma_8$ given is that for  the simulation with massive neutrinos.}
\label{tab:partsimuls}
\end{table*}

\subsection{The numerical simulations}
\label{sec:simdetails}

We primarily used particle-based simulations, summarised in Table \ref{tab:partsimuls}. 
We also ran several simulations using the Fourier-space implementation
of neutrinos using a grid with $512^3$ elements, but as these were not used in 
the final results we omit them from the table for clarity.

Our fiducial simulations have $512^3$ dark matter particles in a box of $150\Mpch$. 
We tested $M_\nu = 0.15$~eV, $0.3$~eV and $0.6$~eV, running $6$ simulations for 
each value of $M_\nu$ with different realisations of the initial
conditions, to suppress cosmic variance. 
To check we had correctly sampled the large scale modes, to explore the linear regime and to compare directly to 
\cite{Viel:2010}, we ran also simulations with box sizes of $512\Mpch$. 
To facilitate comparison with \cite{Brandbyge:2008} and \cite{Viel:2010}, our fiducial cosmology had 
$\Omega_M = 0.3$, a spectral index $n_s = 1.0$, the Hubble parameter $h = 0.7$, and amplitude of the primordial 
power spectrum $A_\mathrm{s} = 2.27 \times 10^{-9}$. This implies that a simulation without neutrinos
has $\sigma_8 = 0.87$ at $z=0$, but a simulation with $M_\nu = 0.3$~eV has $\sigma_8 = 0.8$.
We tested our formula using a variety of simulations with different cosmological parameters, as summarised in Table \ref{tab:partsimuls}.

The initial redshifts of our simulations were $z_i = 99$ for $M_\nu = 0.6$~eV, $z_i = 49$ for $M_\nu = 0.3$~eV 
and $z_i = 24$ for $M_\nu = 0.15$~eV, which we found to be the earliest redshift where the neutrinos were fully non-relativistic. 
Our initial conditions were generated using the Zel'dovich approximation, and
the linear theory power spectrum was calculated by CAMB.
We acknowledge that second order Lagrangian perturbation theory is in principle more accurate, but
our simulations are started at sufficiently high redshift that the effect should be small.
The transfer function used for our cold dark matter (CDM) particles was a weighted average of 
the linear theory transfer functions for CDM and baryons, 
to account for the slight difference between them \citep{Somogyi:2010}. 
Neutrinos and CDM had identical initial random phase 
information, to simulate adiabatic initial conditions.

The numerical convergence of our results and the impact of neutrino shot noise were 
checked with simulations S60ND, S15NU and S60NU.
We further checked the effect of baryonic physics with simulation S60BA, which included $512^3$ baryon particles.
S60BA modelled radiative cooling and star formation using the default multiphase model of \gadget-3
\citep{Springel:2003}. Differences much larger than $10\%$ 
have been found for $k > 20 \hMpc$ between simulations 
implementing different radiative processes 
(e.g. metal cooling) or feedback recipes in $\Lambda$CDM models 
(see e.g. \cite{Rudd:2008, Guillet:2010, vanDaalen:2011}).
Furthermore, the model of AGN feedback suggested by \cite{vanDaalen:2011} 
in order to reproduce observations of groups of galaxies produces 
a power spectrum suppression of roughly $10\%$ at $k = 1 \hMpc$ and $1\%$ at $k = 0.3 \hMpc$.
We did not include this effect; however, \cite{vanDaalen:2011} found it to be
independent of cosmological parameters,  suggesting that
it is also independent of the neutrino mass, and thus should not affect our results.

\subsection[The HALOFIT formula]{The \halofit~formula}
\label{sec:halofit}

\halofit~\citep{Smith:2003} estimates non-linear growth from a given linear theory power spectrum, 
based on the principles of the halo model. These assert that the growth of halos 
should depend only on the local physics at the scale of the halo, 
and not the details of the pre-collapse material, nor the 
larger-scale distribution of matter. 
Thus growth deep in the non-linear regime 
becomes independent of the details of the linear power spectrum, depending 
only on the non-linear scale and the slope and curvature of the power spectrum.
In the mildly non-linear regime, the non-linear correction 
is assumed to be a function of the instantaneous linear power spectrum, without 
explicit dependence on the historical growth rate. 
\halofit~is accurate to $5-10\%$  at $z<3$ \citep{Heitmann:2010a},
suggesting that these approximations are reasonably
accurate for cold dark matter. 

\halofit~realises the above by splitting the non-linear power spectrum into two components. 
A quasilinear term $\Delta^2_\mathrm{Q}$, and a non-linear term $\Delta^2_\mathrm{H}$. 
This means the dimensionless power spectrum can be written as, 
\begin{align}
        \Delta^2 \equiv \frac{k^3 P(k) }{2 \pi^2} = \Delta^2_\mathrm{Q} + \Delta^2_\mathrm{H}\,.
         \label{eq:halofit}
\end{align} 
Both terms depend on the non-linear scale, $k_\sigma$, defined as the wave-number
where the perturbation amplitude, filtered by a Gaussian, becomes greater than unity 
and linear theory breaks down. 

The quasilinear term, $\Delta^2_\mathrm{Q}$, is given as an enhancement to 
the linear power spectrum, and incorporates an exponential cut-off ensuring 
it is negligible on small scales. The form of the quasilinear term is, 
\begin{align}
        \Delta^2_\mathrm{Q}(k)  &= \Delta^2_\mathrm{L}(k) \left[ \frac{(1+ \Delta^2_\mathrm{L} (k))^{\beta(n)} }{1+\alpha(n)\Delta^2_\mathrm{L}(k)} \right] \exp [-f(y) ] \,.\\
	\text{Here}\quad 3 + n &= \left.\frac{d \ln \sigma^2 }{d\ln R} \right|_{k=k_\sigma}\,,
        \label{eq:haloquasilin}
\end{align}
so that $n$ is the effective spectral index of the power spectrum.  
$\beta$ and $\alpha$ are functions of $n$ and $f(y) = y/4 + y^2/8$. 
We refer readers interested in the exact values of these functions to Appendix C of \cite{Smith:2003}. 
$ \Delta^2_\mathrm{L}$ is the dimensionless linear theory power spectrum.

Non-linear growth, $\Delta_\mathrm{H}^2$, dominates when $y \equiv k/k_\sigma \gg 1$. 
The form of  $\Delta_\mathrm{H}$ used by \halofit~is,
\begin{align}
        \Delta_\mathrm{H}^2 (k)  &= \frac{\Delta_\mathrm{H}^{2 \prime} (k)}{1 + \mu(n)/y + \nu (n)/y^2}, \\
        \Delta_\mathrm{H}^{2 \prime} (k)  &= \frac{a(n,\mathcal{C}) y^{3f_1(\Omega)}}{1 + b(n,C) y^{f_2(\Omega)} + [c(n,\mathcal{C}) f_3(\Omega) y]^{3-\gamma(n,\mathcal{C})}} \,.\\
	\text{Where}\quad\mathcal{C} &= \left.\frac{d^2 \ln \sigma^2 (R) }{d\ln R^2} \right|_{k=k_\sigma}\,,
        \label{eq:halononlin}
\end{align}
the spectral curvature of the power spectrum.
$a$, $b$, $c$, $\mu$ and $\nu$ are functions of $n$ and $\mathcal{C}$, and $\nu(n)$ is 
not related to neutrinos. $f_{1,2,3}$ are functions of $\Omega_M$, the total matter density. 
For the values of these functions, we again refer to Appendix C of \cite{Smith:2003}. 
Note that the non-linear growth depends only on $y$, $n$, $\mathcal{C}$ and the matter fraction. 

Massive neutrinos affect this prescription because they suppress the linear power spectrum, 
increasing the non-linear scale and retarding growth. \halofit~estimates 
the effect of this suppression on non-linear power, but neglects any effect of the 
neutrinos on the process of halo collapse, as well as the effects 
of non-linear growth on the neutrinos themselves. 

\section{Results}
\label{sec:results}

\begin{figure*}
\centering
\includegraphics[width=0.45\textwidth]{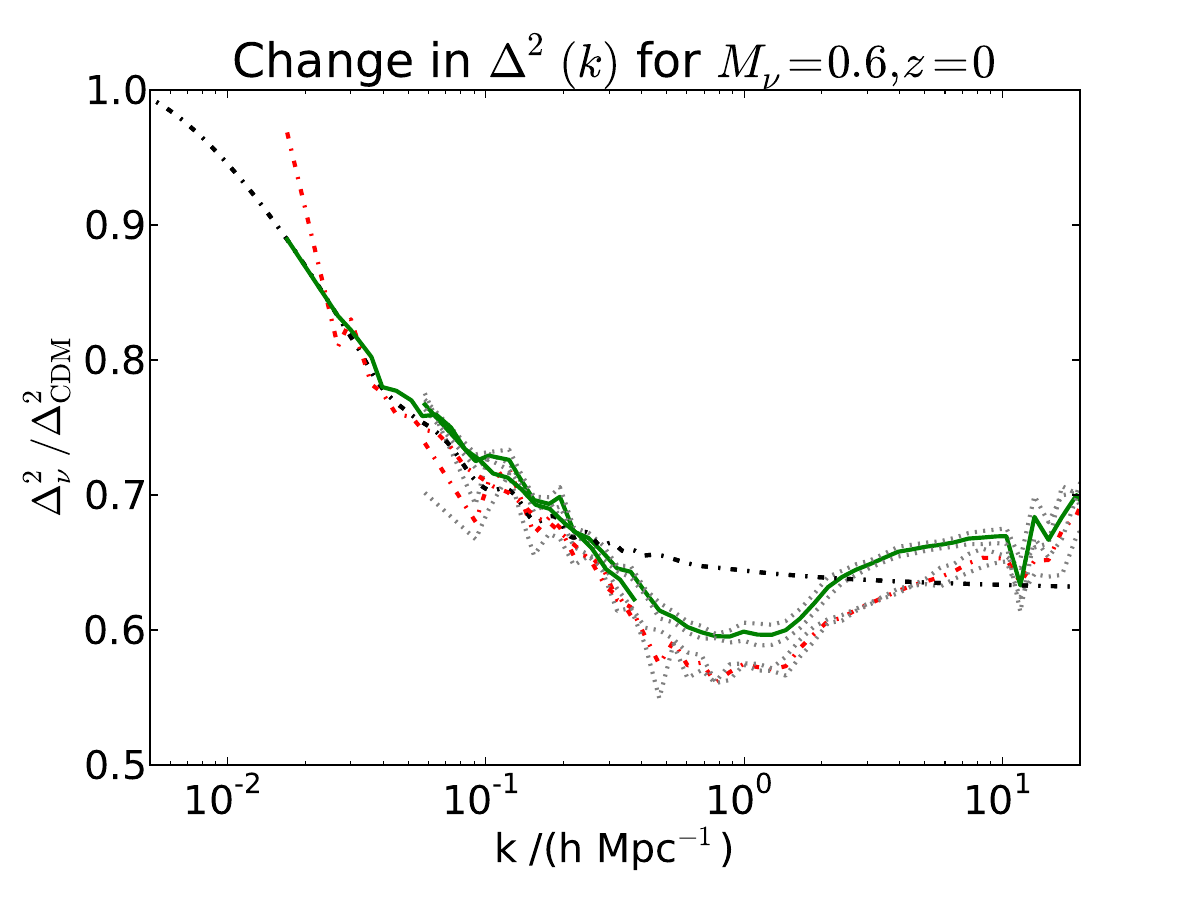}
\includegraphics[width=0.45\textwidth]{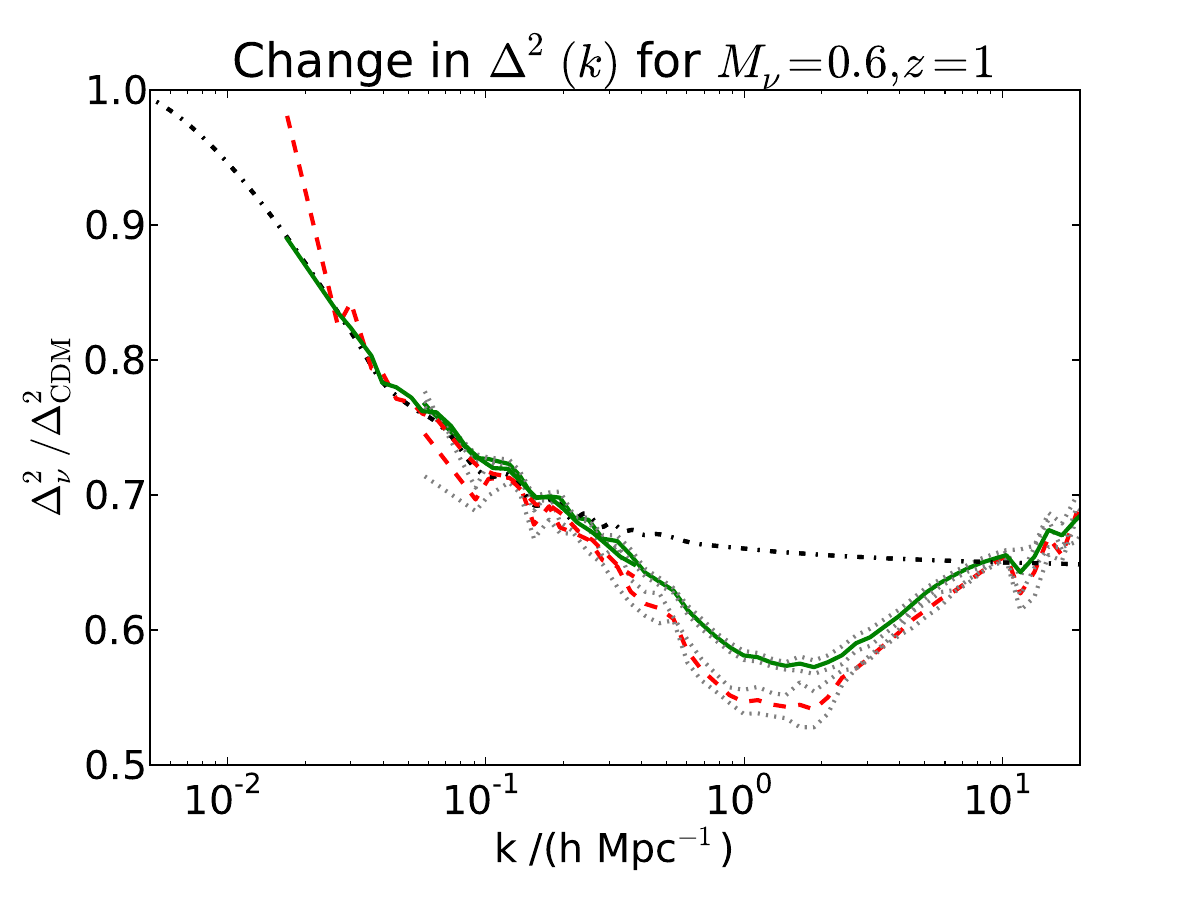}
\caption{The ratio between the matter power spectrum at $z=0$ (Left) and $z=1$ (Right) 
for simulations with and without massive neutrinos ($M_\nu = 0.6$~eV). Solid green lines 
show results for simulations using particles, while dashed red 
lines show results from Fourier-space simulations. The black dot-dashed lines
show the predicted effect from linear theory. 
The simulations used were L60 and S60; box sizes were $512\Mpch$ and $150\Mpch$, with parameters shown in Table \ref{tab:partsimuls}. 
For the smaller box sizes, more than one realisation of structure is available, and 
hence we show (dotted grey lines) the one $\sigma$ error due to sample variance, as
estimated numerically. Particle and Fourier-space methods use the same initial structure realisations. 
The Fourier-space method has more power in the largest scale mode, probably 
because the phase structure of the neutrino component has been prevented from evolving, 
enhancing the effect of sample variance. 
}\label{fig:shape03}
\end{figure*}

\begin{figure*}
\centering
\includegraphics[width=0.45\textwidth]{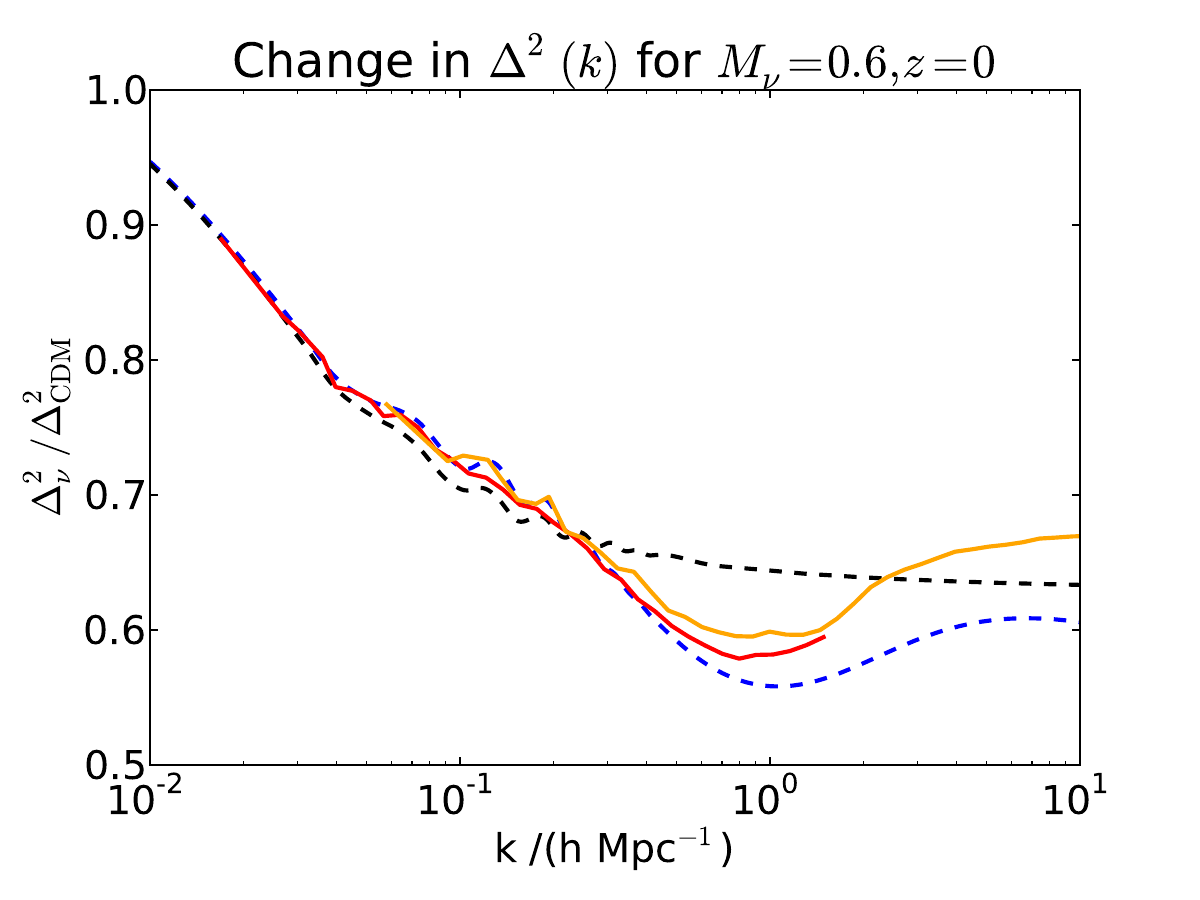}
\includegraphics[width=0.45\textwidth]{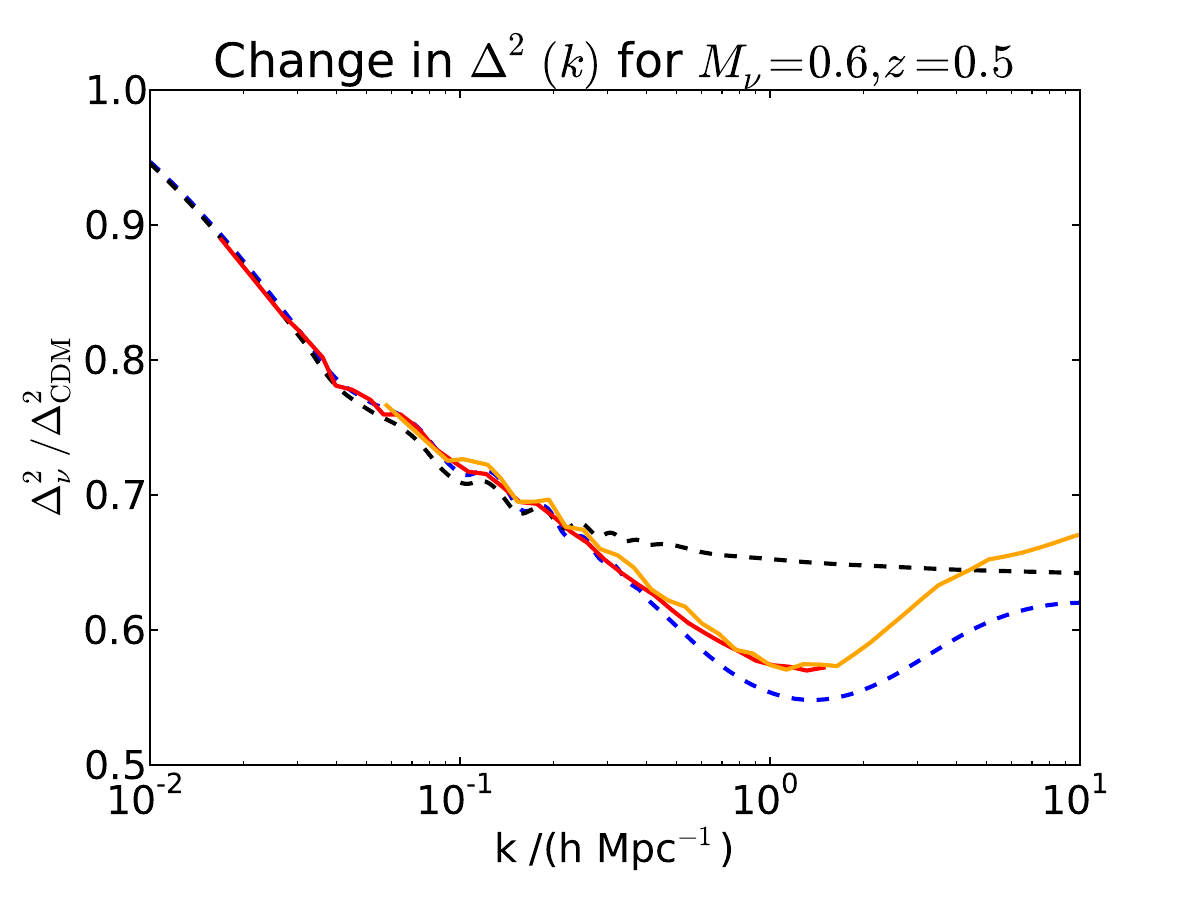}
\includegraphics[width=0.45\textwidth]{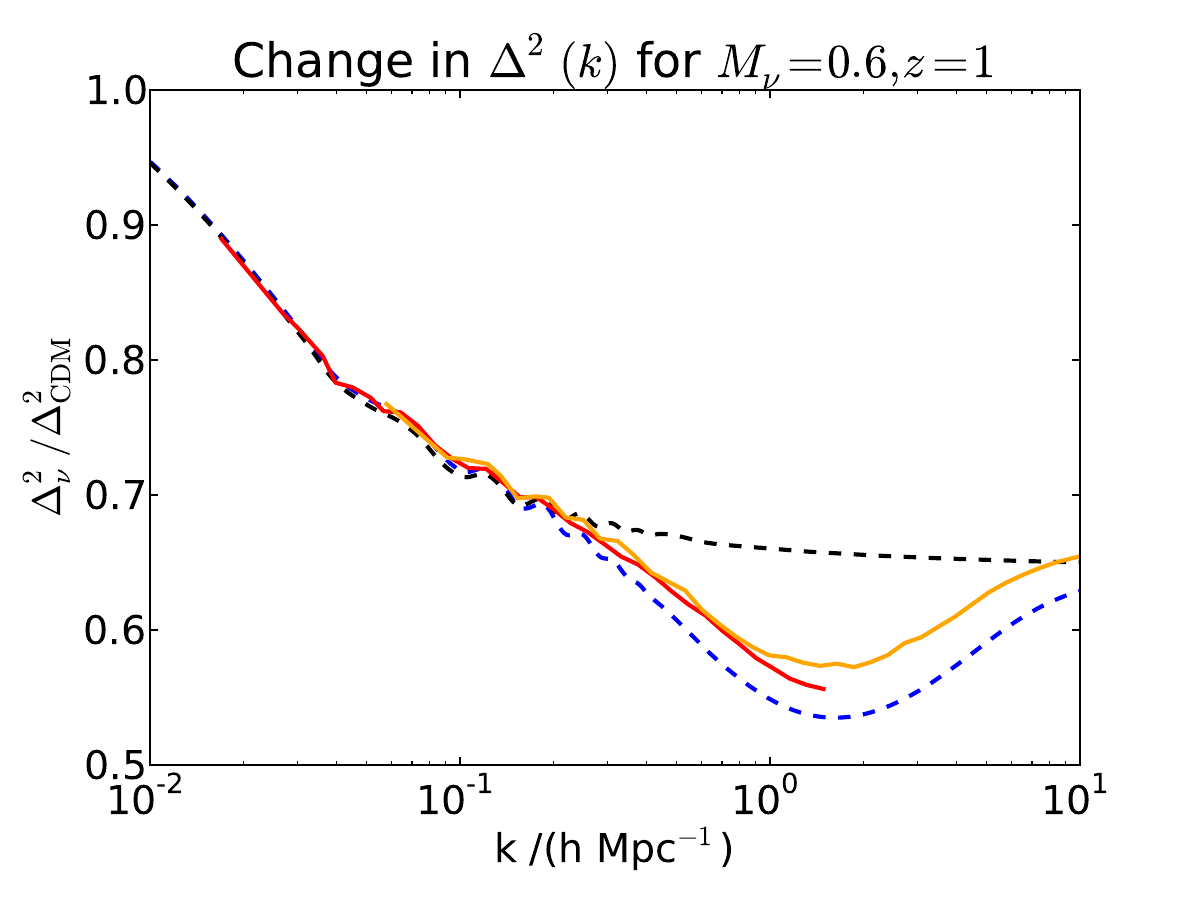} 
\includegraphics[width=0.45\textwidth]{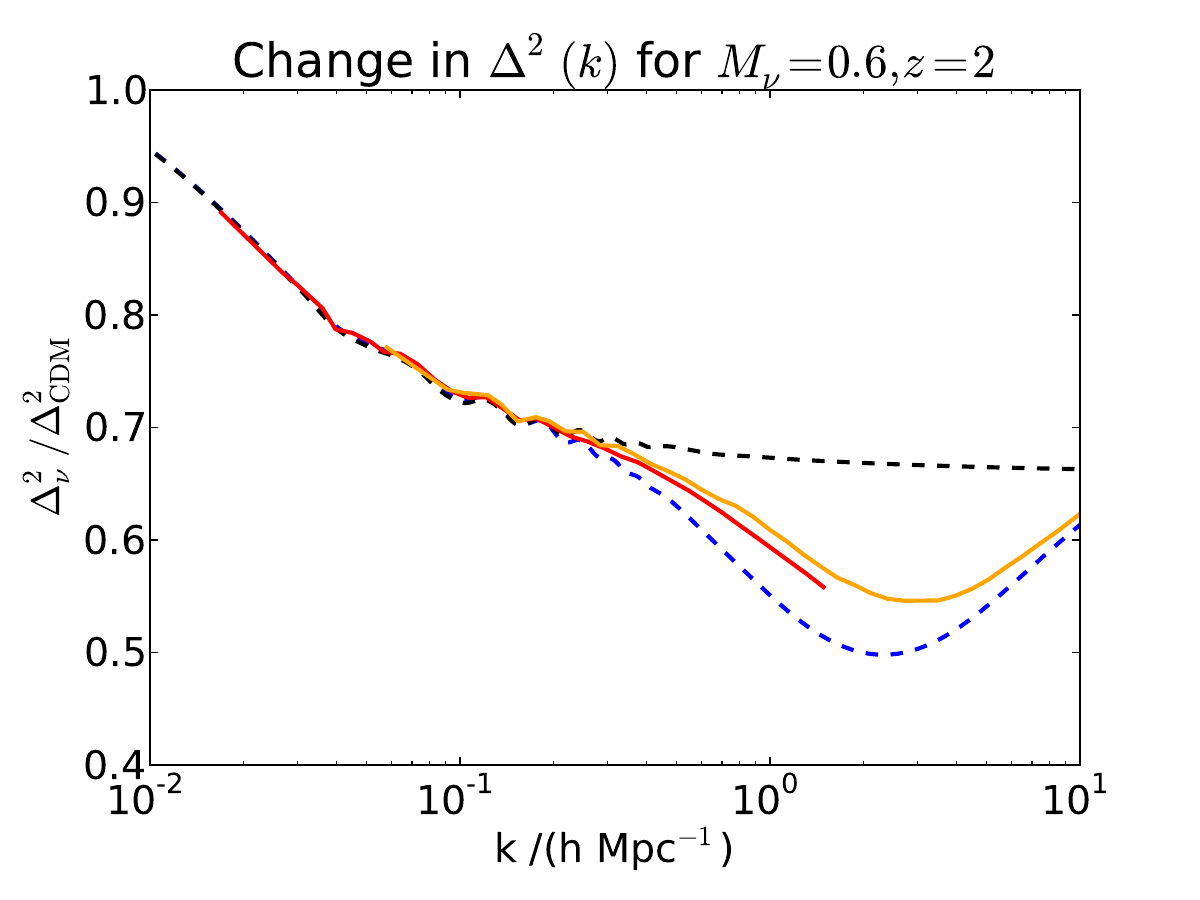}
\caption{The effect of massive neutrinos on the matter power spectrum for a neutrino mass of $M_\nu = 0.6$ eV. 
Solid lines show the ratio between simulations with and without massive neutrinos, for
both L60 (red), with a $512\Mpch$ box and S60 (orange), with a $150\Mpch$ box.  Initial redshift was $99$.
The blue dashed line shows the estimated ratio using \halofit, while the black dashed line shows 
the prediction from linear theory. 
}\label{fig:neutrino06}
\end{figure*}

\begin{figure*}
\centering
\includegraphics[width=0.45\textwidth]{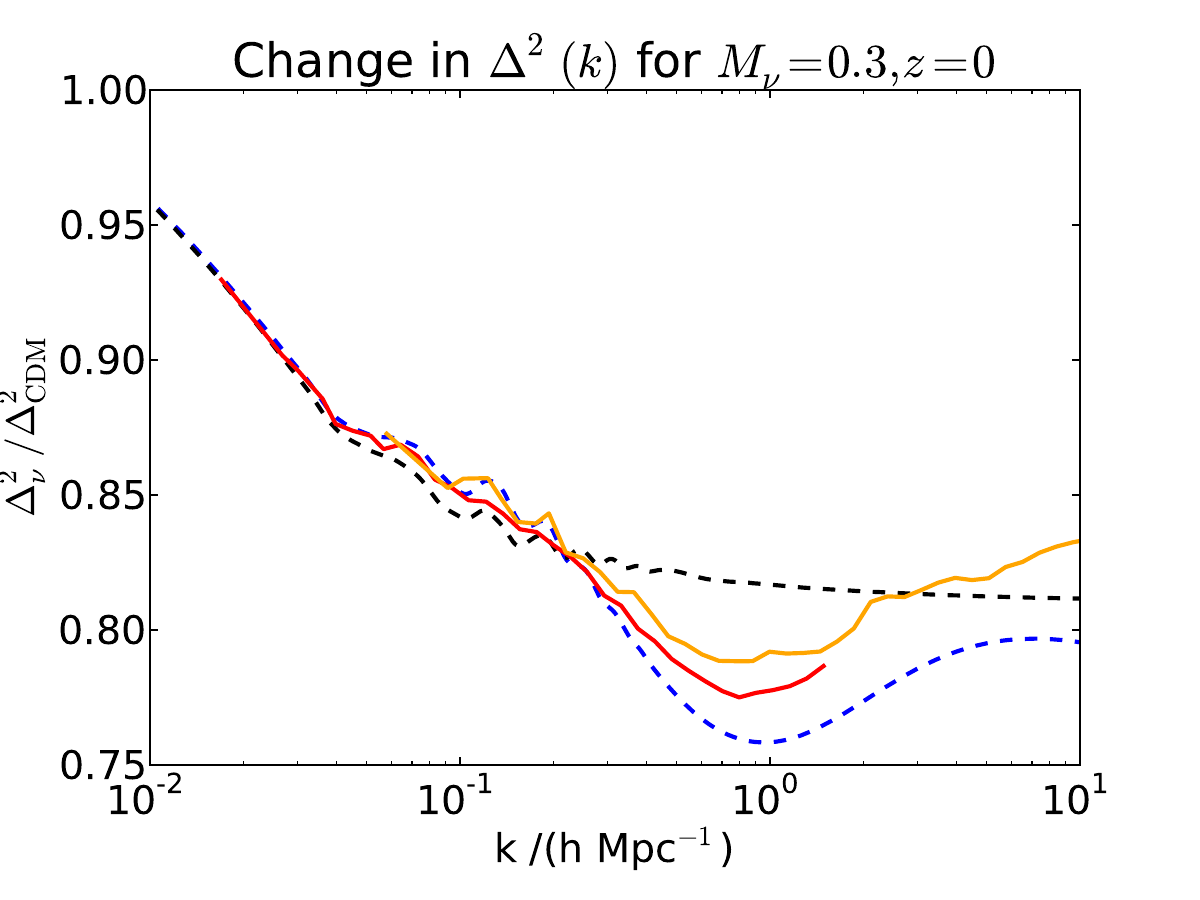}
\includegraphics[width=0.45\textwidth]{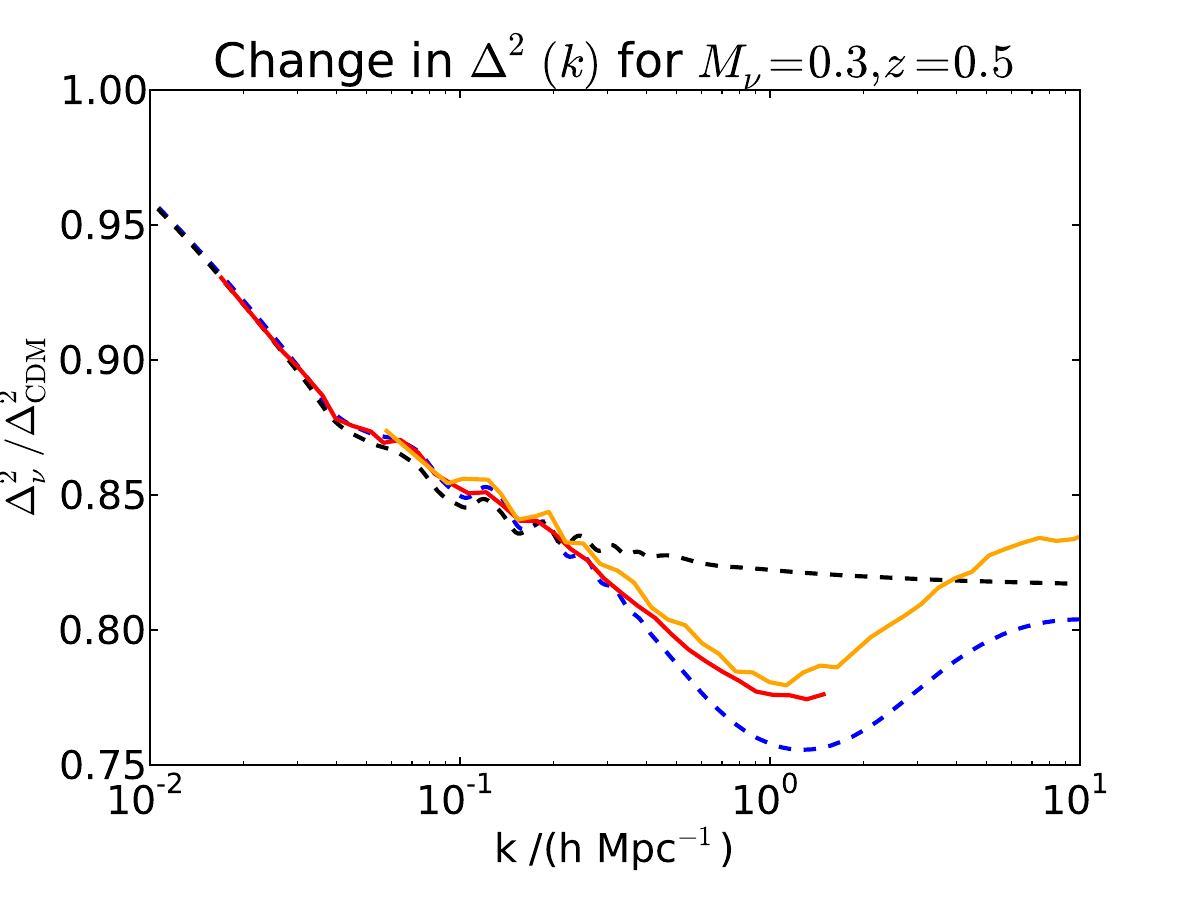}
\includegraphics[width=0.45\textwidth]{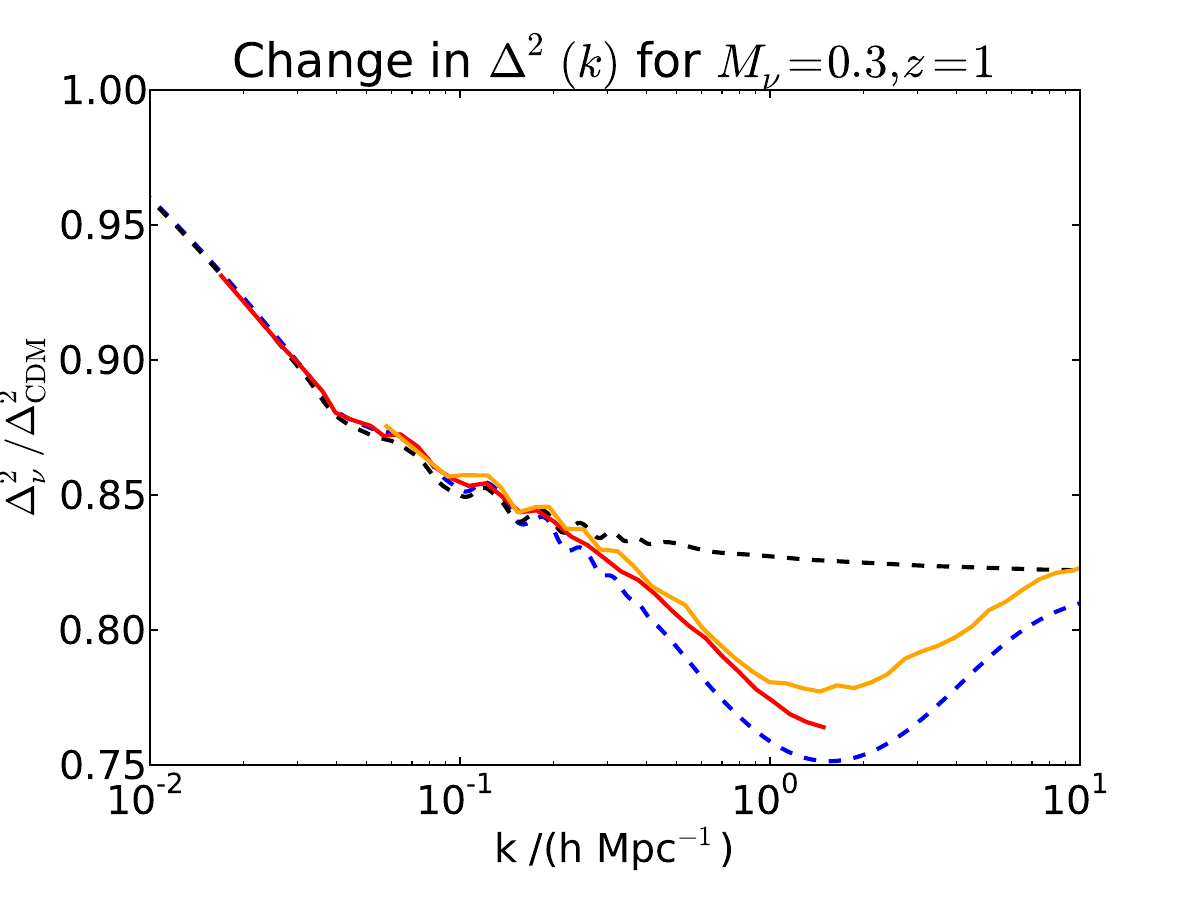} 
\includegraphics[width=0.45\textwidth]{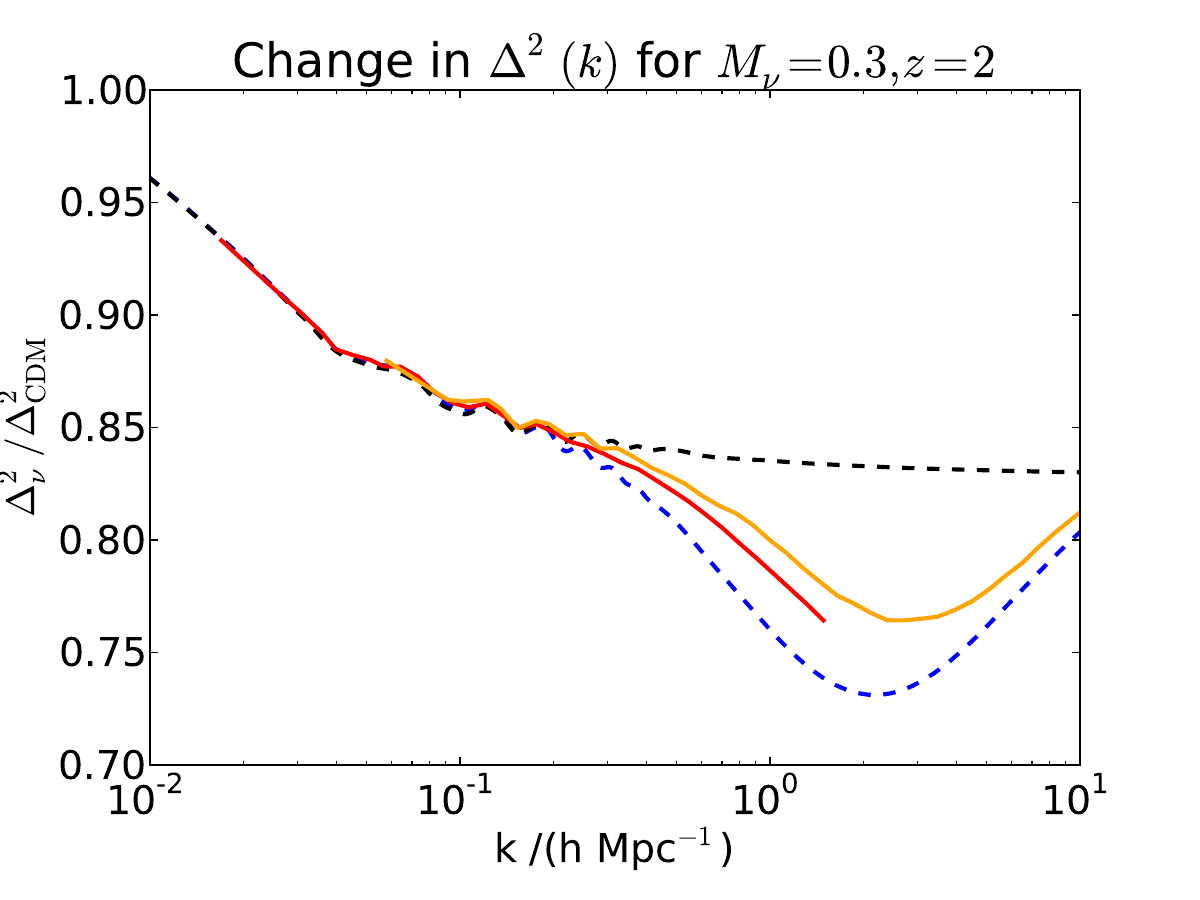}
\caption{The effect of massive neutrinos on the matter power spectrum for a neutrino mass of $M_\nu = 0.3$ eV. 
Solid lines show the ratio between simulations with and without massive neutrinos, for
both L30 (red), with a $512\Mpch$ box and S30 (orange), with a $150\Mpch$ box. Initial redshift was $49$.
The blue dashed line shows the estimated ratio using \halofit, while the black dashed line shows 
the prediction from linear theory. 
}\label{fig:neutrino03}
\end{figure*}

\begin{figure*}
\centering
\includegraphics[width=0.45\textwidth]{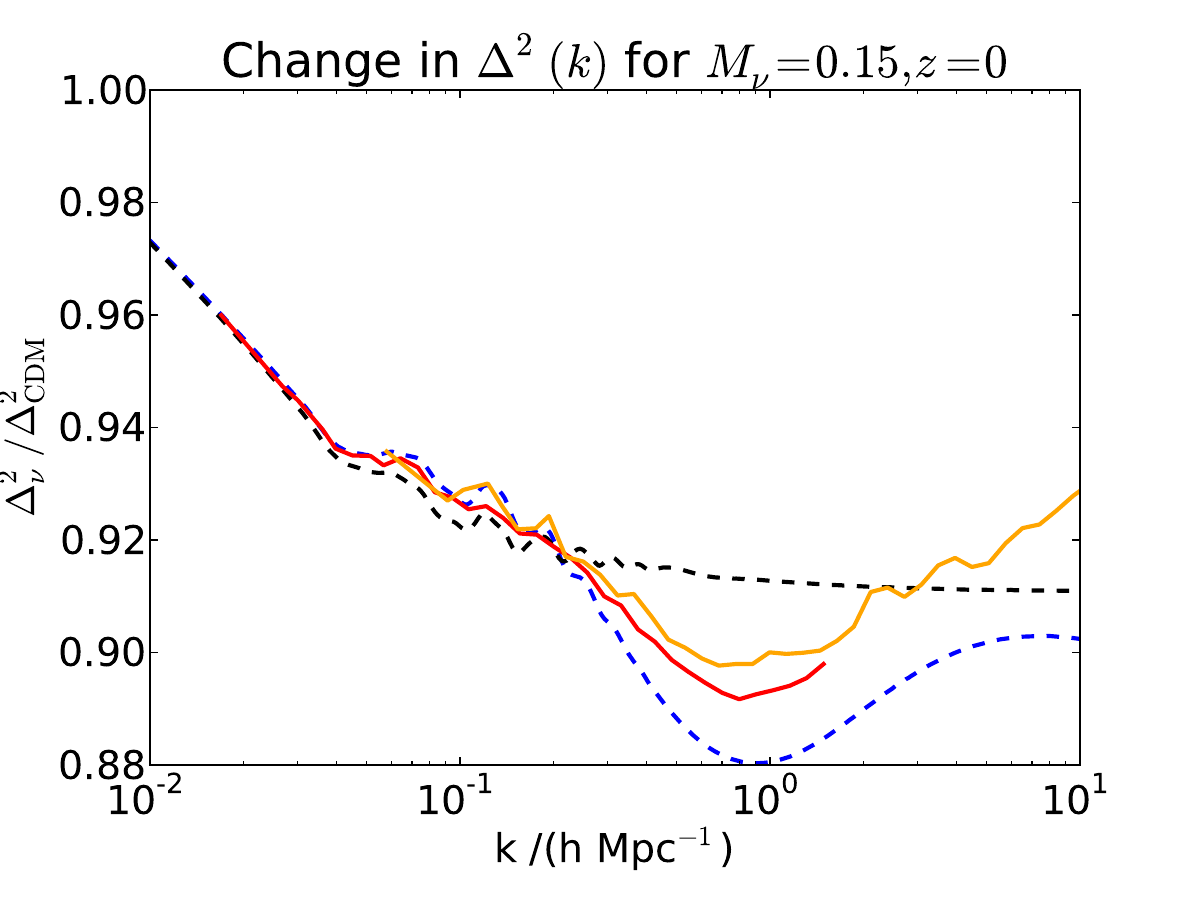}
\includegraphics[width=0.45\textwidth]{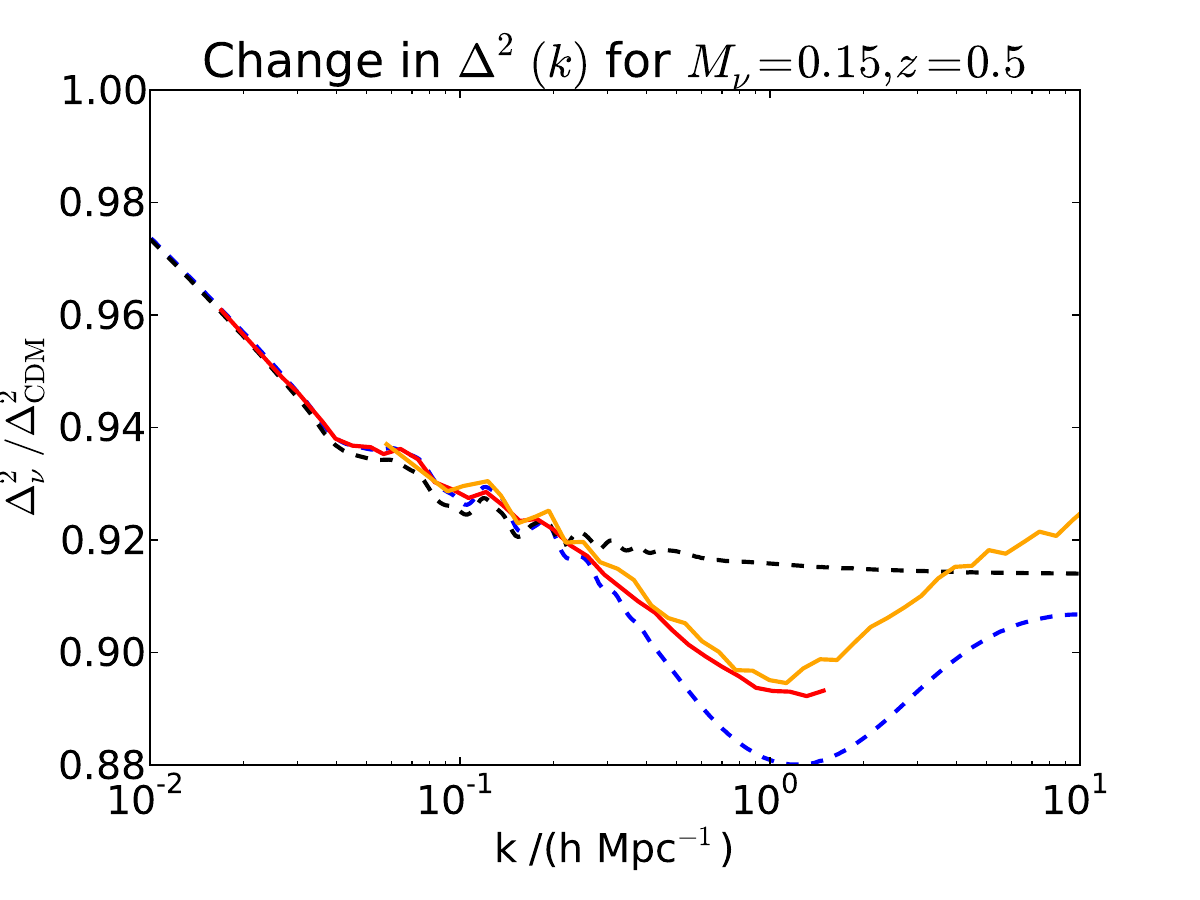} 
\includegraphics[width=0.45\textwidth]{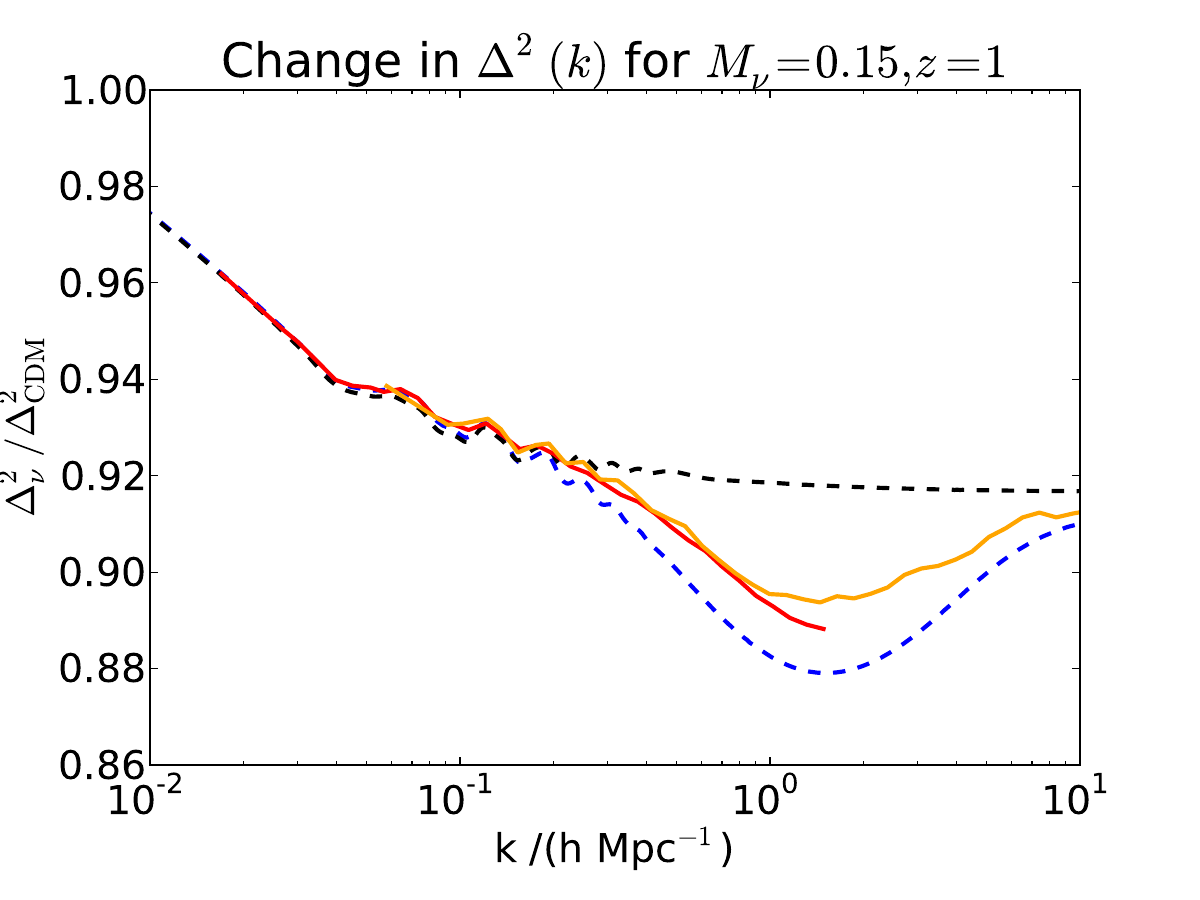} 
\includegraphics[width=0.45\textwidth]{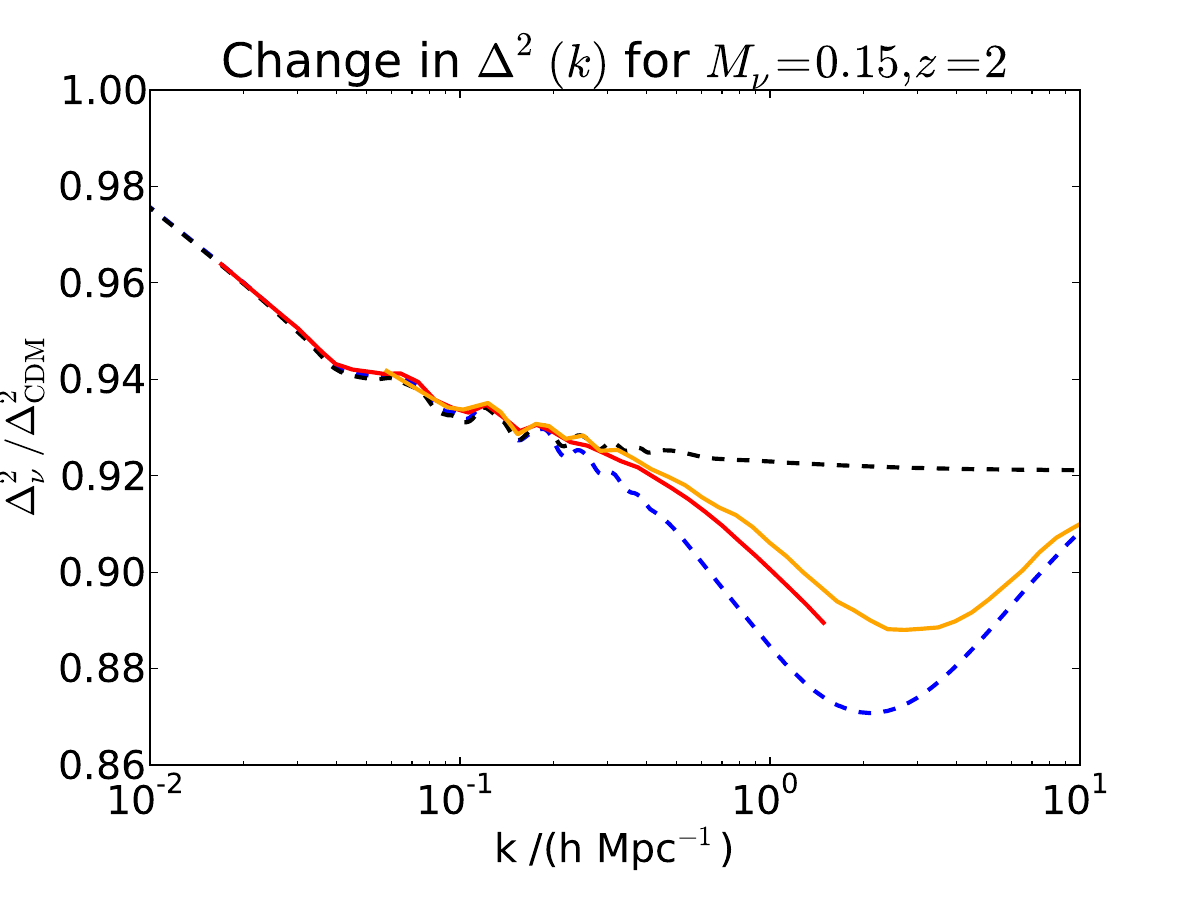}
\caption{The effect of massive neutrinos on the matter power spectrum for a neutrino mass of $M_\nu = 0.15$ eV. 
Solid lines show the ratio between simulations with and without massive neutrinos, for
both L15 (red), with a $512\Mpch$ box and S15 (orange), with a $150\Mpch$ box. Initial redshift was $24$.
The blue dashed line shows the estimated ratio using \halofit, while the black dashed line shows 
the prediction from linear theory. 
}\label{fig:neutrino015}
\end{figure*}

%
Figure \ref{fig:shape03} shows ratios between the matter power spectrum for massive neutrinos, 
and $\Lambda$CDM. Plotted are the predictions of linear theory, of simulations S60 and L60, 
and of grid-based simulations with identical parameters.
The maximal power spectrum suppression in linear theory is proportional to 
$-8 f_\nu$, compared to a maximal non-linear suppression of approximately
$-10 f_\nu$ for $z \leq 1$, and $-10.5 f_\nu$ for $z = 2$.
As discussed in Section \ref{sec:neutrinogrowth}, linear theory predicts an increasing suppression 
approaching an almost flat plateau on small scales.
In the numerical simulations, the flat plateau is replaced by a
minimum. The suppression increases up to  $k \sim 1 \hMpc$, after which it gradually decreases. 
This distinctive spoon shaped pre-virialization feature was 
also found by \cite{Brandbyge:2008} and \cite{Viel:2010}, and in CDM simulations by \cite{Lokas:1996, Smith:2007}. 

We can understand this shape by considering the effect of neutrinos 
on non-linear growth. The suppression caused by massive neutrinos 
delays the onset of non-linear growth and increases the wave-number 
of the non-linear scale. Between the non-linear 
scale for the simulation without neutrinos and the 
non-linear scale for the simulation incorporating neutrinos, the 
suppression will be enhanced by the absence of non-linear growth. 
On the smallest scales, the non-linear growth lost as a consequence 
of the massive neutrinos becomes an ever-smaller fraction of the total, 
and the extra suppression decreases. Eventually non-linear growth 
comes to dominate the linear effect, and the suppression 
is less than predicted by linear theory.

As mentioned in Section \ref{sec:simdetails}, we performed checks for 
numerical convergence. A simulation with increased resolution, 
S60ND (see Table \ref{tab:partsimuls}), had a relative matter power spectrum 
which differed from the fiducal simulation, S60, by $1-2$\% for $k < 7\hMpc$. 
Furthermore, the relative matter power spectra for our two different box sizes, $512$ and $150\Mpch$, 
were always in good agreement on the largest scales probed by the
smaller box, showing that our smaller 
box size was sufficient for  an accurate estimate of the suppression of 
matter power spectrum due to the free-streaming of the neutrinos.
The larger box showed a slightly increased 
suppression on small scales, probably due to limited resolution.
We found that the effect of baryons was, at all redshifts, 
less than $1$\% at $k< 8\hMpc$, gradually increasing on smaller scales, 
in agreement with the results of e.g. \cite{Jing:2006, vanDaalen:2011}. 

Simulations S15NU and S60NU were run with a higher number of neutrino particles,
to check explicitly the effect of shot noise. We found that, although the 
neutrino power spectrum is indeed shot noise dominated on small scales, 
the power on these scales is so small that their impact on the growth of the 
dark matter power spectrum is negligible. Hence shot noise has only 
a minor impact for simulations with a box size of $150\Mpch$, 
at the level of $1\%$ at the relevant scales and redshifts. 

\subsection{Particle and Fourier-space implementation of neutrinos }
\label{sec:partvsgrid}

Figure \ref{fig:shape03} shows that particle and Fourier-space implementations of neutrinos 
give similar results for the matter power spectrum, and both agree
with linear theory on large scales. 
The particle implementation, however, shows a smaller suppression, 
which scales roughly linearly with $M_\nu$,
particularly near the trough of the dip.

This discrepancy has again a simple physical explanation; structure growth in 
the dark matter induces structure growth in the neutrino component, as the less energetic 
neutrinos fall into the gravitational wells created by the dark matter. 
This increases the power in the neutrinos, drawing them closer to the non-linear dark matter power spectrum and 
leading to a reduced neutrino suppression. Essentially this is a backreaction effect, where the 
dark matter drags the neutrinos with it; since it results from the non-linear growth, 
it is not being fully accounted for by linear theory neutrinos.
Note that structure growth in the neutrino component itself remains linear; 
the effect is due to non-linear growth in the dark matter. 
Because this effect is important at the accuracy we wish to achieve, 
our main results (Figures \ref{fig:neutrino06}, \ref{fig:neutrino03} and \ref{fig:neutrino015})
are based on  simulations with the particle implementation of neutrinos. 

\subsection[Comparison to HALOFIT]{Comparison to the \halofit~model}

Figures \ref{fig:neutrino06}, \ref{fig:neutrino03} and \ref{fig:neutrino015}
show the main results of this paper. 
Each Figure shows the suppression of  the matter power spectrum caused by massive
neutrinos at four redshift snapshots ($z= 0, 0.5, 1$ and $2$), compared
to the predicted  effect from \halofit and linear theory. Figure \ref{fig:neutrino06} 
shows $M_\nu = 0.6$~eV, Figure \ref{fig:neutrino03} has $M_\nu = 0.3$~eV and 
finally Figure \ref{fig:neutrino015} has $M_\nu = 0.15$~eV.

\halofit~clearly over-predicts the suppression of the matter power
spectrum due to massive neutrinos in the non-linear regime. 
The cause is similar to that discussed in Section \ref{sec:partvsgrid}; 
\halofit~includes massive neutrinos only through the linear 
theory neutrinos suppression on the non-linear scale, and thus neglects 
any back-reaction from the dark matter. This interpretation is supported by the good agreement of
\halofit with the Fourier-space simulations. 

The largest discrepancy, around $10$\% of the total suppression, occurs at $k\sim 1 \hMpc$.  
The location of the maximal suppression in the numerical simulation moves to larger scales
at lower redshifts, an effect which is again not captured by \halofit, although 
it was present to some extent in our Fourier-space simulations. 
Furthermore, the amplitude of the suppression decreases with redshift, which is 
to be expected from our discussion in \ref{sec:partvsgrid}; as  
non-linear growth occurs in the dark matter more neutrinos fall into the gravity wells.
A cross-over redshift occurs at $z=1$; on very small scales the suppression
here is equal to that of linear theory, while at lower redshifts it is less.
Note also that in the quasilinear regime, $0.05 < k < 0.2\hMpc$, 
the simulations clearly agree much better with \halofit~rather than linear theory 
The dependence of our results on $M_\nu$ is well described
by a linear relation, with the maximal suppression for a given redshift
being proportional to $f_\nu$, although the redshift dependence is more complicated.

\halofit~over-predicts the effect of neutrinos on the smallest scales. 
This is not due to neutrino physics, but is a discrepancy induced because, 
as found by \cite{Hilbert:2009}, \halofit~under-predicts the growth of non-linear power 
for $k > 2 \hMpc$ in a $\Lambda$CDM universe by up to a factor of two. 
We corrected this by re-fitting the \halofit~parameter that controls the 
small-scale power using our $\Lambda$CDM simulations, after which we could 
reproduce the asymptotic neutrino effect. We detail our modifications in Appendix 
\ref{ap:halofit}. Note that \halofit~also has reduced accuracy at
$z=3-4$, with errors of $15-20$\%, compared to $5-10$\% at $z=0-2$.
Because of this, it fails to accurately predict the 
location of the peak non-linear suppression at $z > 2$. 
We did not correct this effect, as our attempts were found to 
negatively impact accuracy at lower redshifts. 

We have also performed some simulations varying the cosmological 
parameters from our fiducial values, as described in Section \ref{sec:simdetails}.  
The results of these simulations were similar 
to those for our fiducial cosmology, and agree with the results 
discussed in \cite{Viel:2010}. The dependence of the non-linear
suppression on cosmology is largely 
captured by \halofit, with the exception of a weak dependence on
$\Omega_M$, which we include in our fitting formula.

\subsection{Towards an improved fitting formula}
\label{sec:fitting}

In the previous Section, we compared our simulations to \halofit. While there was some discrepancy, 
it seemed to capture many of the important physical features. In light of this, 
we have designed an improved fitting formula by modifying \halofit to explicitly account for 
the effect of massive neutrinos. Our fitting formula was derived by performing a series of 
least-squares minimisations of $\mathcal{L} = \Delta^2_\mathrm{sim}/\Delta^2_\mathrm{fit}-1$, where
 $\Delta^2_\mathrm{sim}$ is the matter power spectrum from the simulation and
$\Delta^2_\mathrm{fit}$ is  the equivalent from  
the fitting formula. Both spectra have been rebinned to be evenly spaced in $\log k$. 
To find a minimum, we used Powell's method 
as implemented in SciPy\footnote{\url{http://www.scipy.org}}. To avoid any contamination from sample variance, 
we omitted scales with $k < 6 \frac{2\pi}{\mathrm{Box}}$, where Box is the box size. The redshift bins used
in our fits were $z=0,0.2,0.5,1,2,3$. \halofit~models the effect of 
non-linear growth at $z=4$ rather poorly, and there is some residual shot noise present at these 
redshifts in our neutrino simulations. Furthermore, the only current probe of the matter power spectrum 
at these redshifts is the \Lya forest, which should be examined by modelling the flux power 
directly with full hydro simulations, 
as in \cite{Viel:2010}. Hence we do not attempt to provide a good fit for $z>3$. 
We checked that our fit was not significantly altered by
excluding the simulation results  at $z=3$.

Before addressing the effect of neutrinos, 
we attempted to correct the lack of power shown by \halofit on the smallest scales. 
Eq. \ref{eq:halononlin} shows that the asymptotic behaviour of the non-linear term 
in \halofit is given by
\begin{equation}
        \Delta^2_\mathrm{H} \to \frac{a(n)}{c(n) f_3(\Omega)} y^{3(f_1(\Omega)-1)+\gamma(n)} \sim y^{\gamma(n)}\,.
        \label{eq:haloasymp}
\end{equation}
Thus, to correct the behaviour on small scales we altered $\gamma(n)$, fitting it to 
the $\Lambda$CDM simulations labelled in Table \ref{tab:partsimuls} as 
S60, S60AS, S60NS, S60H and S30OL. We were interested only in the asymptotic behaviour, so we
omitted the results from the large simulation  boxes from the fit. 
We thereby included extremely small scales into the fit; 
down to $k = 30 \hMpc$. The numerical minimisation algorithm proved unstable
when constrained to  $k< 6\hMpc$. We checked that the resulting fit is
a good match to larger scales. 

The largest remaining contribution to the discrepancy between 
\halofit and our simulations is the decrease in the maximal neutrino suppression 
due to neutrinos falling into dark matter halos. This 
is proportional to $f_\nu$ and roughly constant on large scales. To account for it  
we modified the nonlinear term with the ansatz
\begin{align}
        Q_\nu &= \frac{l f_\nu}{1+m k^3}, \\
        (\Delta^\nu_\mathrm{NL})^2 &= \Delta^2_\mathrm{NL} (1+ Q_\nu)\,,
        \label{eq:haloppnon}
\end{align}
with $l$ and $m$ found by minimisation. 
We considered other powers of $k$ in the denominator of $Q_\nu$, as well as additional 
terms and powers of the non-linear scale $k_\sigma$, but none of these 
significantly enhanced the fit. 

To correctly match the location of the maximal suppression 
at higher redshift, $z \geq 2$, we needed to alter the quasilinear term, reflecting 
a leakage of non-linear power from small scales to larger scales caused by the high thermal dispersion 
of the neutrinos. Therefore, we modified Eq. \ref{eq:haloquasilin} as follows,
\begin{align}
        \Delta^2_\mathrm{Q}(k)  &= \Delta^2_\mathrm{L}(k) \left[ \frac{(1+ \tilde{\Delta}^2_\mathrm{L} (k))^{\tilde{\beta}(n)} }{1+\alpha(n)\tilde{\Delta}^2_\mathrm{L}(k)} \right] \exp [-f(y) ], \\
        \tilde{\Delta}^2_\mathrm{L} &= \Delta^2_\mathrm{L} \left(1+\frac{p f_\nu k^2}{1+1.5 k^2}\right), \\
        \tilde{\beta}(n) &= \beta(n) + f_\nu (r + s n^2)\,,
        \label{eq:haloppquasi}
\end{align}
where $p$, $r$ and $s$ were found again by minimisation. These coefficients, as well as $l$ and $m$ were found
by simultaneously fitting to the simulations S15, S30, S60, L15, L30 and L60. For the larger box sizes, we
used only scales with $k< 0.5 \hMpc$, and for the smaller boxes, we used $k < 6 \hMpc$. 

Finally, there was some residual dependence on $\Omega_M$. To account for this, we modified $l$ as
$l \to l - t (\Omega_M - 0.3)$, and fit $t$ separately, based on a  minimisation 
with simulations S30, S30OM and S30OL. The numerical values of the coefficients are shown in 
Appendix \ref{ap:halofit}, together with a summary of the fitting formula.
The reduced error in the fit can be defined as 
\begin{equation}
\mathcal{E} = \sqrt{\frac{\mathcal{L}^2}{N_\mathrm{d}}}\,,
\end{equation}
where $N_\mathrm{d}$ is the number of data points used in the fit. 
We find $\mathcal{E} = 0.007$ with our coefficients, compared to 
$\mathcal{E} = 0.04$ with the unmodifed \halofit.

A rough estimate of the error on our fitting formula, including realisation noise at the $1-\sigma$ level, is given by
\begin{align}
\epsilon &= E(k,z) P(k) \,, \\
E(k,z) &= \frac{\log [ 1+k/k_\sigma(z)]}{1+\log[1+k/k_\sigma(z)]} f_\nu\,,
\label{eq:fiterr}
\end{align}
where $k_\sigma(z)$ is again the non-linear scale.  
Differences between the fitting formula and the simulations change sign 
between $z=0$ and $z=0.2$, suggesting the dominant contribution is realisation noise.
The quoted errors are valid  for $k< 7 \hMpc$, and $z \leq 3$. At
$z>3$ our fitting formula  is somewhat less accurate.
The maximal error over all scales is $E_\mathrm{max} = f_\nu$; 
for $M_\nu = 0.3$, $E_{\mathrm{max}} \sim 2\%$. 

Although they were not included in the fit, we checked that our formula accurately 
reproduced simulations with different cosmologies, chosen to be at the
bounds of the range allowed by current data.
The fitting formula errors for S60AS, S60NS, and S60H were very similar to those found when compared to 
the fiducial cosmology, although there were one or two outliers caused by realisation noise.
Checking against S30OM, which has $\Omega_M= 0.25$, produced somewhat larger errors at the highest 
redshift bin, with $z=3$. The lower value of $\Omega_M$ means that $z=3$ is deeper into the quasilinear regime, 
where \halofit~is less accurate. 

\begin{figure*}
\centering
\includegraphics[width=0.45\textwidth]{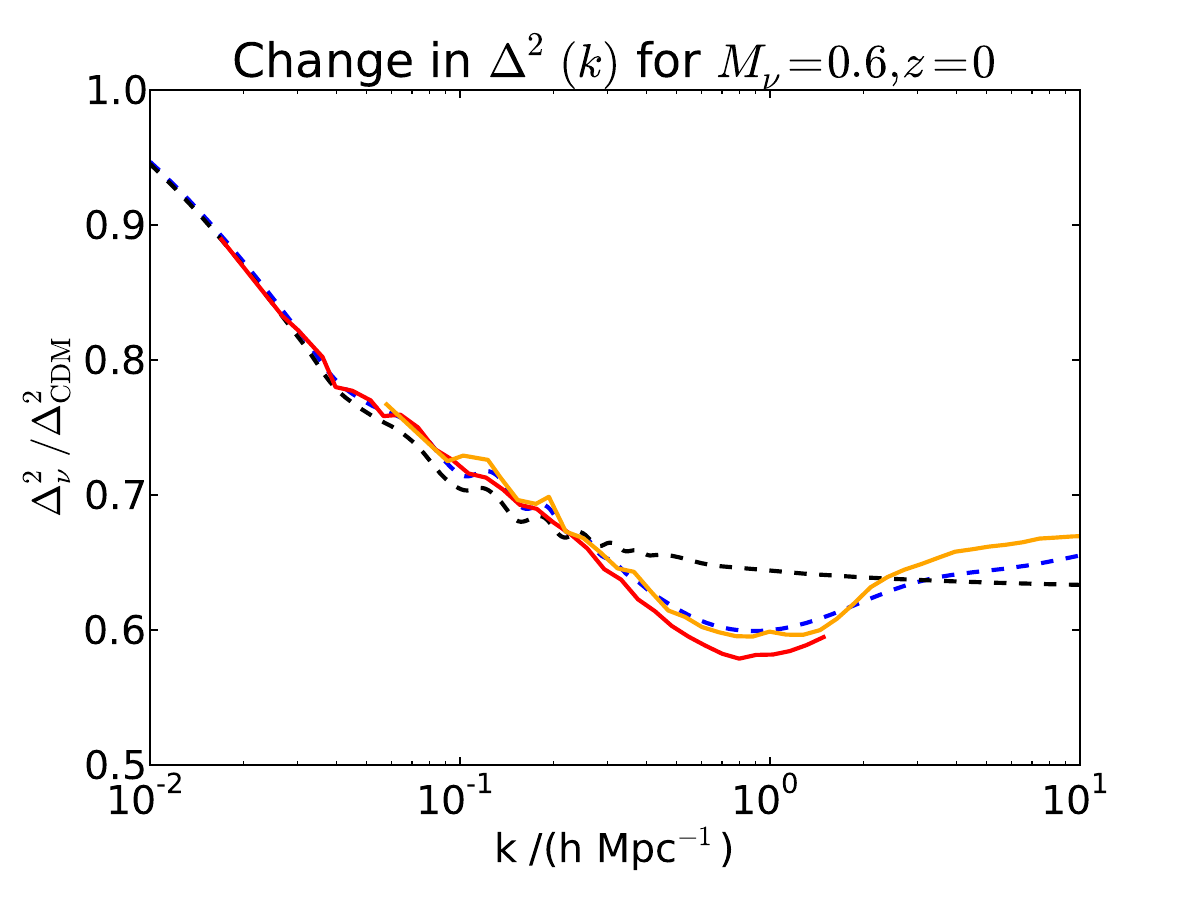}
\includegraphics[width=0.45\textwidth]{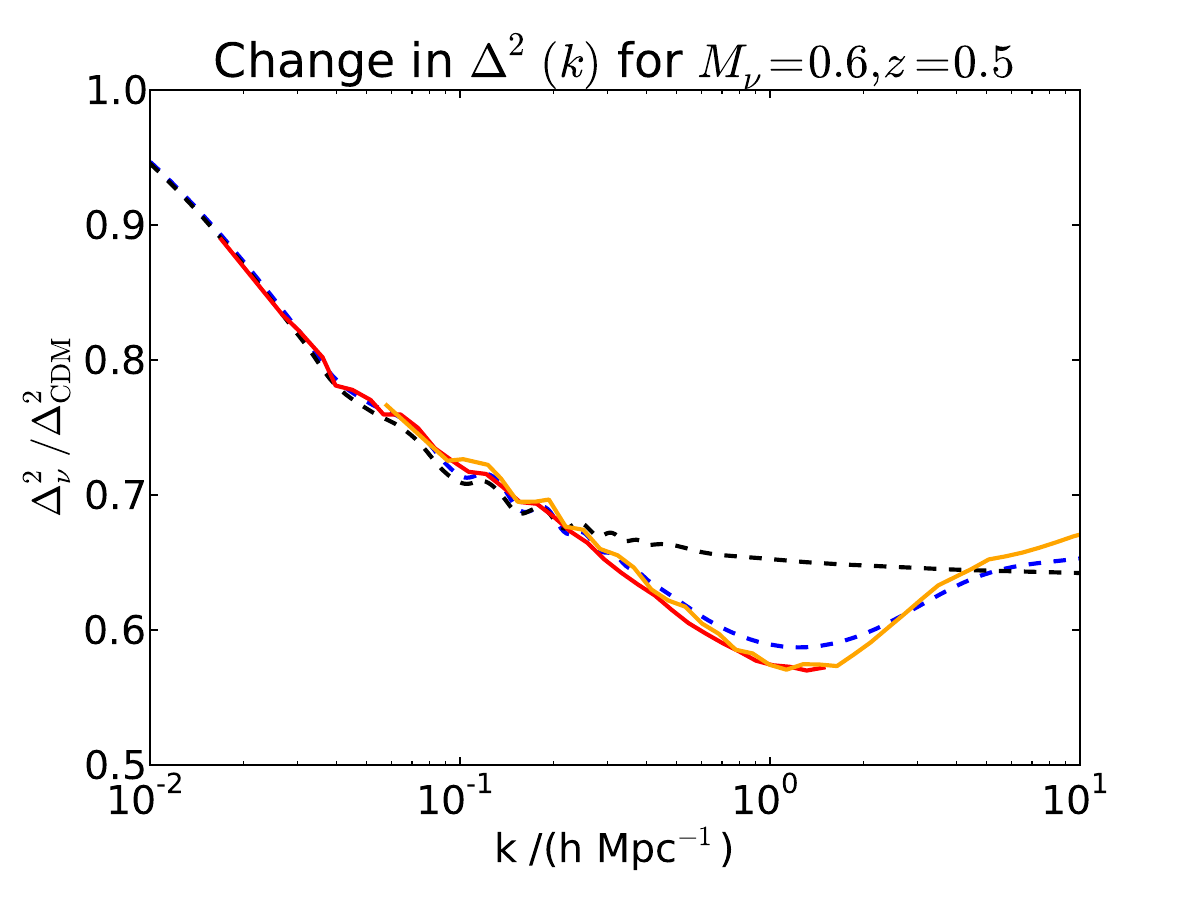}
\includegraphics[width=0.45\textwidth]{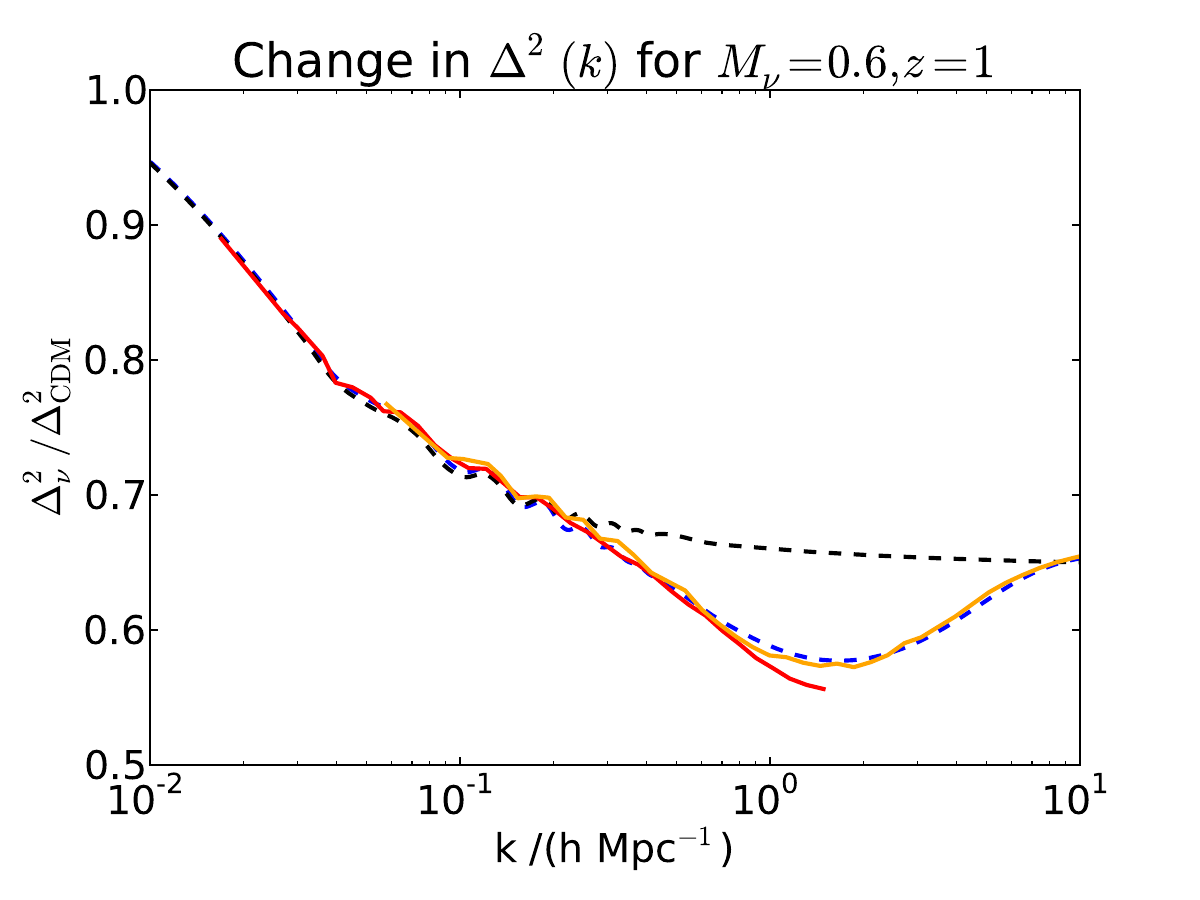} 
\includegraphics[width=0.45\textwidth]{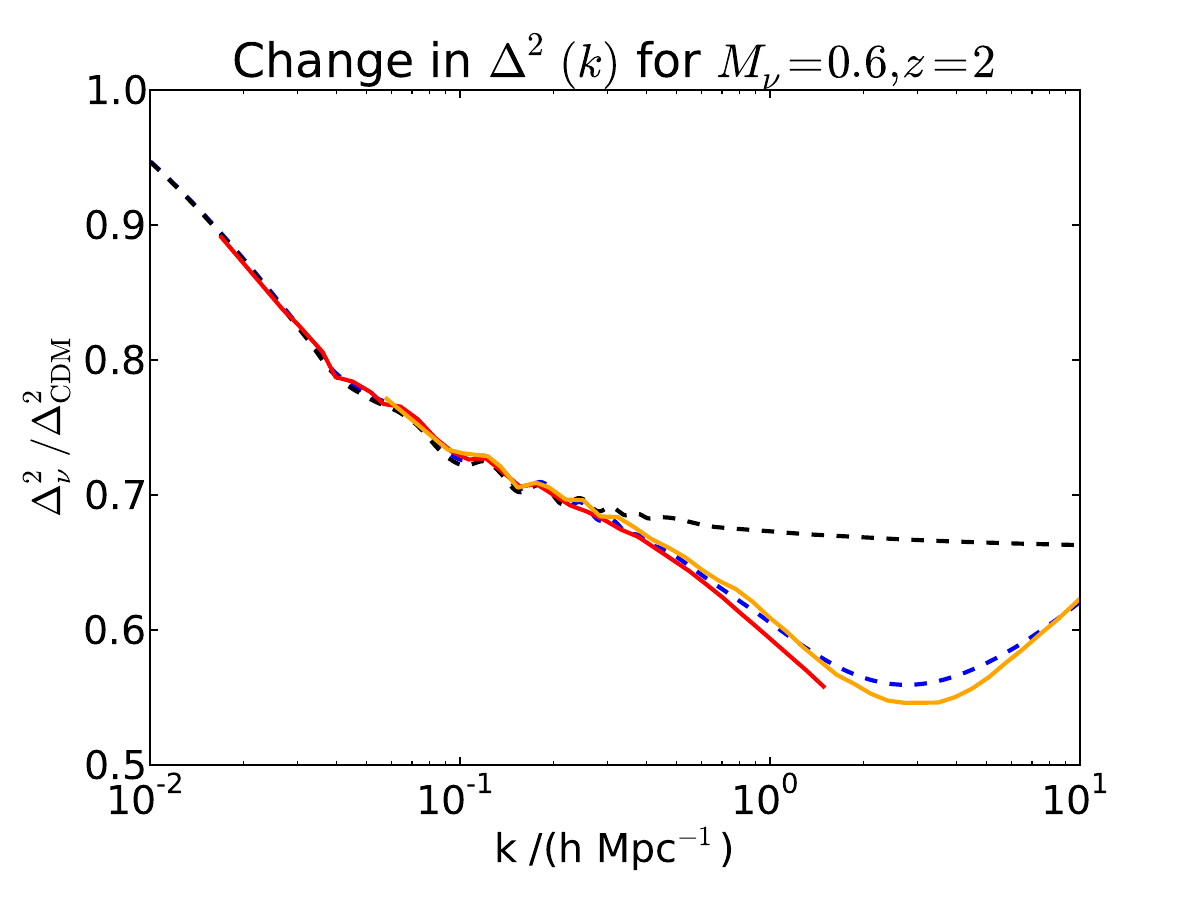}
\includegraphics[width=0.45\textwidth]{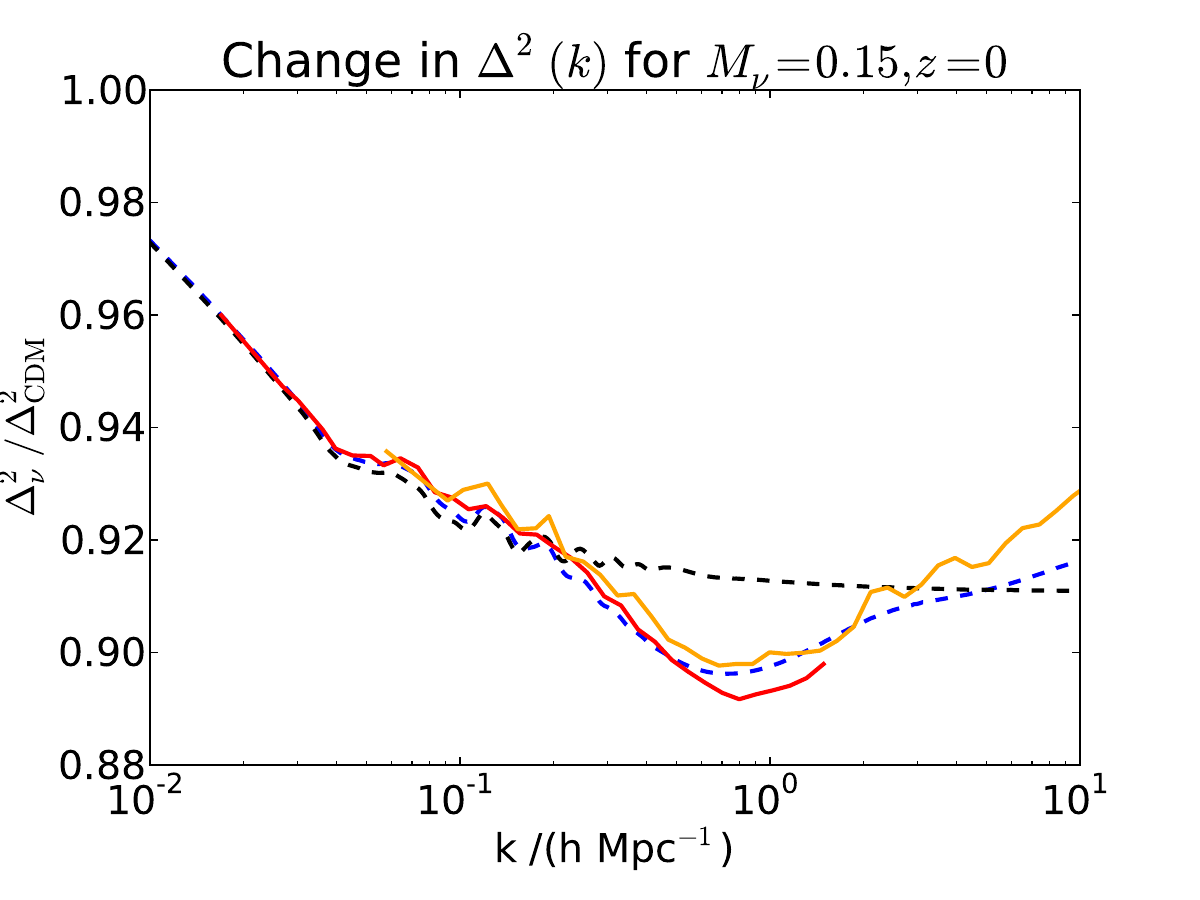}
\includegraphics[width=0.45\textwidth]{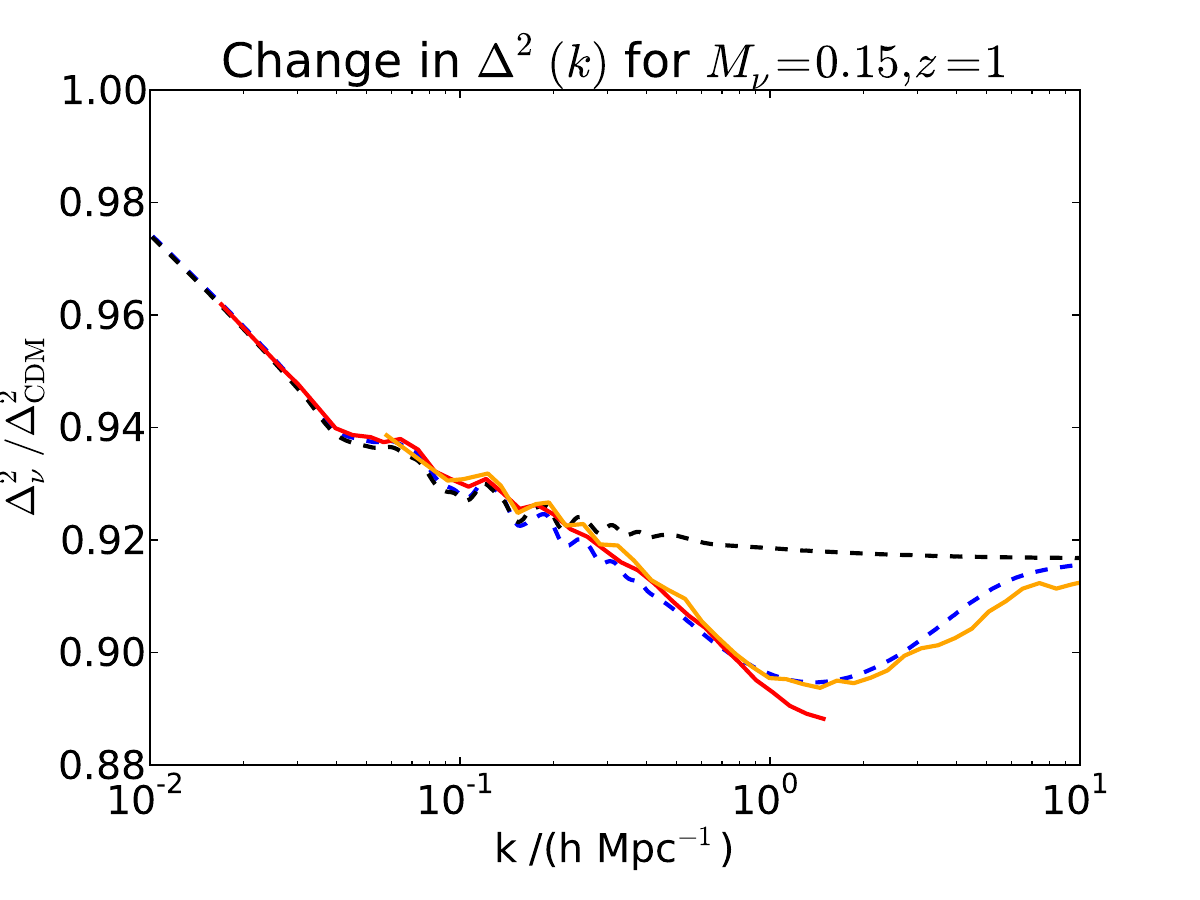} 
\caption{The effect of massive neutrinos on the matter power spectrum for a variety of neutrino masses. 
Solid lines show the ratio between simulations with and without massive neutrinos, for
both $512\Mpch$ (red) and $150\Mpch$ (orange) boxes.
The blue dashed line shows the estimated ratio using our improved fitting formula, 
while the black dashed line shows 
the prediction from linear theory. 
}\label{fig:neutrinofit}
\end{figure*}

Figure \ref{fig:neutrinofit} shows the performance of our improved fitting formula, in a similar format to that of
Figure \ref{fig:neutrino06}. It is clear that our improved formula captures the suppression due to 
neutrinos significantly  more accurately than \halofit, correctly predicting 
the location and depth of the maximal suppression to within the quoted
errors. 

\section{Discussion and Conclusions}
\label{sec:discussion}
 
We have performed a suite of detailed simulations of the matter power spectrum, 
incorporating the effects of massive neutrinos using both particle and Fourier-space methods, 
but focussing on  simulations with neutrino particles. We have compared our results to those 
predicted by the widely used \halofit~approximation to the non-linear
matter power spectrum, and presented an improved fitting formula. 

Observational constraints on the neutrino mass prior to this work have either 
used linear theory \citep{Reid:2010} or \halofit~to estimate the
effect of massive neutrinos on the matter power spectrum.
\cite{Thomas:2009}, for example, used photometric redshifts at
$z=0.45-0.65$, on scales $k \lesssim 0.2 \hMpc$.
This is in the weakly non-linear regime, where we find \halofit~to be accurate. 
Hence our results should not significantly change their published
constraints and that of studies probing similar scales. 

However, while \halofit~reproduces the suppresion of the matter power
spectrum due to the free-streaming  of neutrinos 
well on some scales, in the fully non-linear regime it 
systematically over-predicts the level of the effect.
Furthermore,  \halofit~fails to capture the redshift dependence of the
amplitude and scale  of the peak suppression. Our quantitative
numerical study of the non-linear regime  will thus 
certainly be of importance for future galaxy \citep{Wang:2005,Schlegel:2009,Carbone:2011}, 
CMB lensing \citep{Cooray:1999,Kaplinghat:2003}, \Lya forest \citep{Gratton:2007} and weak lensing \citep{Ichiki:2009,Euclid:2010} surveys.
\cite{Abazajian:2011} have published a review of the potential constraints on the neutrino mass 
from a variety of observational measurements, many of which reach the $M_\nu \sim 0.1$~eV range. 
Realising these limits will require theoretical predictions of the power spectrum accurate to 
the percent level, significantly exceeding the accuracy of current prescriptions. 

We have proposed an improved fitting formula which modifies
\halofit~to take account of the effects of massive neutrinos for
$k < 7\,\hMpc$ at  $z \leq 3$.  Errors are proportional to
the amplitude of the suppression, and thus to $f_\nu$; 
for $M_\nu = 0.3~\mathrm{eV}$, the largest error is roughly $2\%$ of the matter power spectrum.
The maximal non-linear suppression we find is around $-10 f_\nu$,
significantly  greater than  the $-8 f_\nu$ expected from linear theory. Hence constraints obtained
using our formula should 
be tighter than predicted by forecasts using linear theory.

\cite{Saito:2008, Saito:2009, Saito:2011} and \cite{Wong:2008} have proposed an approximation to the non-linear
matter power spectrum  based on one-loop perturbation
theory, while \cite{Lesgourgues:2009}  have modelled the effects of neutrinos with an approach 
based on a time renormalisation group flow. 
These approaches are to some extent complementary to our efforts; while these formulae are designed 
for the quasilinear regime, for $k< 0.3\hMpc$ or $z > 1$, our formula
should be useful even in the fully 
non-linear regime, at the lowest redshifts. 

Our fitting formula should be sufficiently accurate to make forecasts for
the accuracy of neutrino mass measurements of future surveys. 
We should stress, however that we have concentrated here on the
relative impact of the neutrinos; \halofit itself only predicts  
the non-linear power spectrum to an accuracy of $5-10\%$
\citep{Heitmann:2010a} with larger discrepancies at very small scales 
and high redshifts ($z>3$). Detailed quantitative analysis of actual data
reaching into the non-linear regime will thus still need careful calibration with 
accurate  numerical simulations. These will have to be tailored to the 
redshift and scales relevant for the particular data set at hand as the 
simulations still struggle to provide the required dynamical range 
and the fitting formula are not yet accurate at the percent
level over the full range of scales, cosmological parameters, neutrino
masses and redshifts of interest. 

We have implemented our fitting formula in the form of a series of patches to CAMB, 
which have been included in the publicly released version.


\section*{Acknowledgments}

Calculations for this paper were performed on the COSMOS Consortium
supercomputer within the DiRAC Facility jointly funded by STFC, the
Large Facilities Capital Fund of BIS and the University of Cambridge,
as well as the Darwin Supercomputer of the University of Cambridge
High Performance Computing Service (http://www.hpc.cam.ac.uk/),
provided by Dell Inc.  using Strategic Research Infrastructure Funding
from the Higher Education Funding Council for England. We thank Volker
Springel for giving us permission to use \gadget-3.  SB and MH are  supported
by STFC.  MV is partly supported by: ASI/AAE theory grant, INFN-PD51
grant, PRIN-MIUR, PRIN-INAF, the ``cosmoIGM'' ERC starting
grant, and acknowledges support by the European Commission's FP7 Marie Curie
Initial Training Network CosmoComp (PITN-GA-2009-238356).

\appendix

\section[HALOFIT-NU: an improved fitting formula]{\halofit-$\nu$: an improved fitting formula}
\label{ap:halofit}

The \halofit~non-linear term is given by
\begin{align}
        \Delta_\mathrm{H}^2 (k)  &= \frac{\Delta_\mathrm{H}^{2 \prime} (k)}{1 + \mu(n)/y + \nu (n)/y^2} \\
        \Delta_\mathrm{H}^{2 \prime} (k)  &= \frac{a(n,C) y^{3f_1(\Omega)}}{1 + b(n,C) y^{f_2(\Omega)} + [c(n,C) f_3(\Omega) y]^{3-\gamma(n,C)}} \,.\\
	\text{Here}\quad 3 + n &= \left.\frac{d \ln \sigma^2 }{d\ln R} \right|_{k=k_\sigma} 
        \text{and}\quad C = - \left. \frac{d^2 \ln \sigma^2 }{d\ln R^2} \right|_{k=k_\sigma}\,.
\end{align}
To allow \halofit to more accurately model the growth of non-linear power on the smallest scales, we 
modify it by altering $\gamma(n, C)$ to account for the extra growth on small scales as
\begin{align}
        \gamma &\to \gamma+\gamma'  \\
        \gamma' &= 0.316 -0.0765 n -0.835 C
\end{align}
To account for the effect of non-linear growth in the neutrino component, the non-linear term is modified as
\begin{align}
        Q_\nu &= \frac{f_\nu ( 2.080 - 12.4[\Omega_M - 0.3])}{1+1.20\times 10^{-3} y^3} \\
        (\Delta^\nu_\mathrm{H})^2 &= \Delta^2_\mathrm{H} (1+ Q_\nu)\,,
        \label{eq:haloppnonnumeric}
\end{align}
incorporating the residual dependence on $\Omega_M$. 
Finally, the altered quasilinear term is
\begin{align}
        \Delta^2_\mathrm{Q}(k)  &= \Delta^2_\mathrm{L}(k) \left[ \frac{(1+ \tilde{\Delta}^2_\mathrm{L} (k))^{\tilde{\beta}(n)} }{1+\alpha(n)\tilde{\Delta}^2_\mathrm{L}(k)} \right] \exp [-f(y) ] \\
        \tilde{\Delta}_\mathrm{L}^2 &= \Delta^2_\mathrm{L} \left(1+\frac{26.3 f_\nu k^2}{1+1.5 k^2}\right) \\
        \tilde{\beta}(n) &= \beta(n) + f_\nu (-6.49 + 1.44 n^2)\,,
        \label{eq:haloppquasinumeric}
\end{align}
completing our fitting formula.

\bibliography{halo_neutrino}

\label{lastpage}
\end{document}